\documentclass[usenatbib,iop]{aeb_emulateapj_2010}
\usepackage{amsmath}
\usepackage{amssymb}

\usepackage{natbib}

\usepackage[usenames,dvips]{color}

\newcommand{\alf}{Alfv\'en }

\def\Ms{{\rm M}\s} 

\def\Hz{{\rm Hz}} 
\def\GHz{{\rm GHz}} 

\def\m{{\rm m}} 
\def\cm{{\rm c}\m} 
\def\pc{{\rm pc}} 
\def\kpc{{\rm k}\pc} 
\def\Mpc{{\rm M}\pc} 

\def\Ms{M_\odot} 

\def\erg{{\rm erg}} 

\def\Jy{{\rm Jy}} 
\def\mJy{{\rm m}\Jy} 

\def\Sr{{\rm sr}} 

\def\mas{{\rm mas}} 
\def\rad{{\rm rad}} 

\def\cf{{cf.}} 

\def\Ms{M_\odot}
\renewcommand{\d}{{\rm d}}
\newcommand{\e}{{\rm e}}
\newcommand{\sgn}{{\rm sgn}}
\newcommand{\RM}{{\rm RM}}
\newcommand{\RMo}{{\rm RM}_{\rm obs}}
\newcommand\bmath[1] {\mbox{\boldmath$\rm #1$}}
\newcommand{\un}{u_{\rm nth}}
\newcommand{\HARM}{{\small HARM~}}

\def\sinc{{\rm sinc}}

\def\Rfms{R_{\rm fms}}
\def\Rmax{R_{\max}}
\def\Rsim{R_{\rm sim}}

\def\VSOPII{{\em VSOP-2}}

\begin{document}

\title{Parsec-scale Faraday Rotation Measures\\ from General Relativistic MHD Simulations of Active Galactic Nuclei Jets}

\author{
Avery E.~Broderick\altaffilmark{1}
and
Jonathan C.~McKinney\altaffilmark{2}
}
\altaffiltext{1}{Canadian Institute for Theoretical Astrophysics, 60 St.~George St., Toronto, ON M5S 3H8, Canada; aeb@cita.utoronto.ca}
\altaffiltext{2}{Department of Physics and Kavli Institute for Particle Astrophysics and Cosmology, Stanford University, Stanford, CA 94305-4060, USA; Chandra Fellow; jmckinne@stanford.edu}

\shorttitle{pc-scale Rotation Measures from GRMHD Simulations of AGN Jets}
\shortauthors{Broderick \& McKinney}


\begin{abstract}
For the first time it has become possible to compare global 3D general
relativistic magnetohydrodynamic (GRMHD) jet formation simulations
directly to very-long baseline interferometric multi-frequency
polarization observations of the $\pc$-scale structure of active galactic
nucleus (AGN) jets.  Unlike the jet emission, which requires post hoc
modeling of the non-thermal electrons, the Faraday rotation measures
($\RM$s) depend primarily upon simulated quantities and thus provide a
robust way in which to confront simulations with observations.  We
compute $\RM$ distributions of 3D global GRMHD jet formation simulations,
with which we explore the dependence upon model and observational
parameters, emphasizing the signatures of structures generic to the
theory of MHD jets.  With typical parameters, we find that it is
possible to reproduce the observed magnitudes and many of the
structures found in AGN jet $\RM$s, including the presence of transverse
$\RM$ gradients.  In our simulations the $\RM$s are generated within a
smooth extension of the jet itself, containing ordered toroidally
dominated magnetic fields.  This results in a particular bilateral
morphology that is unlikely to arise due to Faraday rotation in
distant foreground clouds.  However, critical to efforts to probe the
Faraday screen will be resolving the transverse jet structure.
Therefore, the $\RM$s of radio cores may not be reliable indicators of
the properties of the rotating medium.  Finally, we are able to
constrain the particle content of the jet, finding that at $\pc$-scales
AGN jets are electromagnetically dominated, with roughly 2\% of the
comoving energy in nonthermal leptons and much less in baryons.
\end{abstract}


\maketitle

\section{Introduction} \label{I}

Spatially resolved radio images of the polarized emission of active galactic
nucleus (AGN) jets on $\pc$-scales, obtained via very-long baseline
interferometry (VLBI), have provided a wealth of
information regarding the large-scale structure of jet magnetic fields
within the emission region.  For many objects these
observations suggest the presence of large-scale, radially-correlated,
toroidally dominant magnetic fields within the emission region
\citep{Udom-Tayl-Pear-Robe:97,List-Mars-Gear:98,gpc00,push05,Jors_etal:05}, in
broad qualitative agreement with simple analytical models of
relativistic magnetized outflows \citep[e.g.,][]{lb02a,lb02b,lpg05,zbb08}.

Performing these observations often requires removing the rotation of
the polarization angle, $\Psi$, associated with Faraday rotation, a
consequence of propagation through intervening, non-radiating plasma.
If the entirety of the Faraday rotation occurs in the foreground, ignoring
finite beam effects, the observed polarization angle at a wavelength,
$\lambda$, is related to the intrinsic polarization angle via,
\begin{equation}
\Psi_{\rm obs} = \Psi_{\rm int} + \RM\lambda^2\,,
\end{equation}
where the Faraday rotation measure, $\RM$, is independent of wavelength and
has a value that depends upon the plasma parameters along a given line
of sight.  In practice, eliminating the Faraday rotation requires
imaging the source at a number of wavelengths, accurately registering
each image to an absolute reference frame, verifying at each
independent position within the image that the polarization angles
follow the expected $\lambda^2$ law and fitting to obtain a map of the
$\RM$s across the image.  Despite the difficulty in performing these
measurements, $\pc$-scale $\RM$ maps for a number of blazars and BL Lac
objects can now be found in the literature
\citep{rcg01,zn01,Zava-Tayl:01,asada02,Zava-Tayl:02,Zava-Tayl:03,gc03,cdfffss03,Zava-Tayl:04,aiuk04,Gabu-Murr-Cron:04,zt05,Asad_etal:08a,Asad_etal:08b,ainkn08,gmjar08,OSul-Gabu:09a,kgosb09,cog10}.

Typical values of the $\RM$s of AGN jets range from roughly
$1\,\rad\,\m^{-2}$ to $10^4\,\rad\,\m^{-2}$, with extreme core
$\RM$s of $10^5\,\rad\,\m^{-2}$ being observed in a handful of
sources (though these are generally at the centers of cooling core
clusters).  While the measuring of $\RM$s allows the elimination of
a foreground effect upon the structure of the polarization of AGN
jets, it also presents an opportunity to study the intervening,
non-radiating plasma.  The source of the $\RM$s is not a priori
clear, with potential contributions arising from propagation through
the Galactic interstellar medium, intracluster medium, the galactic
halo of the AGN, the AGN broad-line region and through the circum-jet
environment.  It is this last possibility, propagation through the
plasma immediately surrounding the jet, that is the subject of this
paper.

Free-free absorption
from $\pc$-scale foreground clouds has been detected in a handful of
sources, implying that foreground clouds exist and may be responsible
for a significant fraction of the observed $\RM$s in some systems
\citep[e.g., Cen A and NGC,][]{Jone_etal:96,Walk_etal:00}, though
there is no reason to expect these to result in ordered $\RM$ maps
exhibiting gradients that are correlated with other aspects of the jet
structure.  In contrast, there are a number of reasons to think that
the circum-jet environment
is responsible for substantial fraction of the AGN jet $\RM$s.
Firstly, there is considerable structure on the scale of the resolved
jet, which is unlikely to arise from Galactic or intergalactic
propagation.  Second, in many sources there is a clear gradient
in the $\RM$ values along the jet axis, steeply decreasing with
increasing distance from the radio core.  Finally, in a handful of objects
\citep[e.g., 3C 273,][]{Zava-Tayl:01,asada02,Gabu-Murr-Cron:04,zt05,Asad_etal:08a,Asad_etal:08b}
there is now strong evidence for transverse $\RM$ gradients, which
have thus far been interpreted in the context of a circum-jet Faraday
screen containing large-scale toroidal magnetic fields
\citep{blandford93,gpc00,Brod-Loeb:09b}.

Magnetic tower models are often invoked to provide a reasonable
basis for the presence of helical fields in relativistic jets.
However, models identified as magnetic tower types in the literature
are non-relativistic models \citep{kc85,lynden96,nuh01,llkur02} and
thus are not directly applicable to astrophysical jets \citep{lpg05}.
Nevertheless, non-relativistic simulations of magnetic tower models
suggest that the $\RM$s are strongly influenced by the toroidal
magnetic field, and correlated with the large-scale jet structure
\citep{nuh01,kunhc04,ukhnc04,lk05,ntll08}.  Comparisons between
idealized, qualitative magnetohydrodynamic (MHD) models \citep{lccbp06,lcb06}
and observations of $\kpc$-scale polarized emission from FRI jets
suggest that in addition to an ordered toroidal magnetic field, at
these distances such jets must also contain a smaller, though still
significant, disordered poloidal field.  However, at such large
distances FRI jets are strongly affected by their interaction with the
external medium, and thus these results may have little to do with the
$\pc$-scale, intrinsic structure of AGN jets in general.

Self-consistent quantitative models of idealized relativistic MHD jets
have been obtained more recently using a combination of analytical
modelling \citep{vk04} and time-dependent simulations
\citep{Komi-Bark-Vlah-Koni:07,tmn09,tnm09}.  However, such idealized
simulations cannot consider the self-consistent formation of the jet
by the black hole interacting with an accretion disk or the effects of
magnetic field geometry and turbulence within the disk.  Idealized
studies cannot, e.g., self-consistently determine the effects of black
hole spin on the jet power for a given mass accretion rate.
In contrast, modern general relativistic magnetohydrodynamical (GRMHD)
simulations of accreting black holes allow a study of the self-consistent
generation of relativistic jets from rotating, accreting black holes
\citep{McKi:06,hk06,bhk08,mb09}.  GRMHD simulations allow one to study
the conditions under which relativistic jets are produced, including the
role of the magnetic field geometry near the black hole and how MHD
turbulence affects the formation of the jet from the black hole
or the wind from the turbulent disk
\citep{mg04,bhk08,mb09}.  Such simulations find that an ordered (but
not necessarily large-scale) dipole-like magnetic field within the
accretion disk leads to a quasi-steady relativistic outflow
powered by the Blandford-Znajek mechanism \citep{bz77,mg04}.  While
such jets are dominated by a toroidal magnetic field, they remain
stable against the disruptive helical kink (screw) mode due to relativistic rotation
of magnetic field lines, sideways (lateral) expansion,
finite mass-loading, and non-linear saturation \citep{mb09}.

Synchrotron emission maps have been produced for highly-idealized
relativistic magnetized jet models
\citep{lpg05,zbb08,gvatb09}, but they generally have major
simplifying assumptions, such as perfect cylindrical geometry,
self-similarity, only solving for the asymptotic structure
(which is unconstrained relative to the self-consistent global solution
that would include the region near the disk),
treating the accretion disk as a boundary condition, or ignoring
gravity altogether.  As far as the authors are aware, none of the
advanced, but still somewhat idealized,
relativistic MHD jet simulations \citep{Komi-Bark-Vlah-Koni:07,tmn09,tnm09}
have been used to study the emission from AGN jets.
Moreover, in light of the
lack of an ab initio understanding of particle acceleration
within AGN jets, all efforts to model the jet emission require
ad hoc prescriptions for inserting the emitting particles.

Unlike emission maps, the $\RM$ maps depend only upon the velocity,
magnetic field, thermal ion density and electron temperature.  As a
consequence MHD simulations are well-suited to computing $\RM$s without
additional assumptions.  Already, global axisymmetric MHD simulations
have been used to constrain the mass accretion rate and magnetic field
geometry of the supermassive black hole in the Galactic center
\citep{sqs07}.  Computing the $\RM$s of AGN jets requires the
self-consistent simulating of both the emitting fast jet core and the
surrounding Faraday screen.  However, this is naturally accomplished
by GRMHD simulations of jet formation, which generally produce a
variety of outflow components, including an ultra-relativistic jet and
a surrounding trans-relativistic wind \citep{McKi:06}.

Here we present the first analyses of $\pc$-scale $\RM$s associated
with self-consistent 3D GRMHD simulations of relativistic jets.
For this purpose, we employ the fully 3D GRMHD simulation reported by \citet{mb09}, and
perform polarized radiative transfer through the 3D volume, generating
frequency-dependent polarized flux maps and $\RM$s.  We investigate
how these depend upon model, observational and physical parameters,
paying specific attention to how Faraday rotation within the circum-jet plasma
may be distinguished from that due to distant Faraday screens.
Section \ref{sec:CM} summarizes the computational modeling, including
the GRMHD simulation, radiative transfer and beam convolution
computations.  Section \ref{sec:Implications} presents $\RM$ maps
for a variety of model parameters and describes the implications for
the location and structure of Faraday screen, as well as our
theoretical understanding of jet formation.  Finally, our conclusions
are summarized in Section \ref{sec:C}.

\section{Computational Methods} \label{sec:CM}

A number of factors affect the computation of the $\RM$ maps of
simulated jets.  Foremost among these is the location, structure and
dynamical state of the Faraday screen itself, which we will find
to be naturally generated by the circum-jet material immediately surrounding the jet.
However, in addition we must also carefully
account for the relativistic motion within the Faraday screen, the
suppression of Faraday rotation at high electron temperatures, finite
radio-beam effects and even the location and mechanism responsible for
producing the observed jet emission.  In this section we describe how
we do each.

\subsection{GRMHD Simulations} \label{sec:CM-NS}

To model the jet system, we use the fully 3D (no assumed symmetries)
general relativistic (GRMHD) simulations reported in \citet{mb09}.
We will only summarize the aspects that are relevant
to our present application here.  The simulations were generated using
the conservative, shock-capturing GRMHD scheme \HARM \citep{gmt03} to simulate the evolution of an
accretion flow around a rotating black hole, and the subsequent
self-consistent generation of bipolar relativistic outflows.
Given an initially dipole magnetic field, embedded within
the accretion disk near the black hole with spin $a/M\approx 0.92$,
they have found that generally a quasi-stable relativistic jet
develops and propagates to large radii.
While in general many different black hole spins could be considered,
models with $a/M\gtrsim 0.4$ are qualitatively similar
to those with $a/M\approx 0.92$ \citep{mck05}.
The simulations run for a duration of about $5\times10^3GM/c^3$,
by which the jet has passed beyond the simulation box, which has a
radius of $10^3GM/c^2$.
The baryon density at the base of the jet, very close to the rotating black hole,
is determined by limiting the comoving electromagnetic energy density per rest-mass energy density
to no larger than $100$, which strictly limits the Lorentz factor to less than order $100$.
The accretion disk has a finite extent of about $r\sim 40M$,
which limits the radial range (order $r\sim 100M$) over which the jet is collimated by the
disk, corona, and disk-driven wind.
This finite collimation extent leads to a saturation
of the Lorentz factor at roughly 5--10.
The effect of a more or less extended disk, and so larger or small terminal Lorentz factors,
respectively, can be considered in future work.
The jet near the black hole is perturbed by disk-driven turbulence
that induces substructure dominated by the $|m|=1$ (where $m$ is the
azimuthal quantum number) mode at any given radius,
and there is also substantial power in other spherical harmonics.
Varying amplitudes of these modes are advected with the jet velocity
and become part of the jet substructure at large radii.
A single snapshot at $t=2750GM/c^3$ was used from the simulation data,
which is a time corresponding to after the jet has fully developed inside the computational box
but the flow has not significantly reflected off the box.

The simulations are completely described by the gas rest-mass density,
$\rho$, the gas internal energy, $u_g$ (gas pressure
  $p_g=\left(\Gamma_g-1\right)u_g$), the lab-frame velocities
$\bmath{\beta}$, the comoving magnetic fields, $\bmath{b}$, and the
spacetime geometry (defined by the metric components, $g_{\mu\nu}$).
While these simulations assume an ideal gas
equation of state, with a fixed adiabatic index of $\Gamma_g=4/3$, variations
in the adiabatic index do not significantly change the dynamics of
highly magnetized jets \citep{mg04}.
From these quantities one can compute the baryon number density,
$n\equiv\rho/m_b$ (where $m_b$ is the baryon mass)
and the ideal gas temperature, $T=p_g/nk$ (where $k$ is Boltzmann's constant).
Since we are interested in rapidly accreting, and therefore well-coupled,
systems, we assume that the electron and ion temperatures are the
same.  Finally, note that the lab-frame and comoving magnetic fields
are related by
\begin{equation}
\bmath{B}
=
\Gamma \left( \bmath{1} - \bmath{\beta}\bmath{\beta}\right)\cdot\bmath{b}
\quad\Leftrightarrow\quad
\bmath{b}
=
\frac{1}{\Gamma} \left( \bmath{1} + \Gamma^2 \bmath{\beta}\bmath{\beta}\right)\cdot\bmath{B}
\,,
\label{eq:Bb}
\end{equation}
where $\Gamma\equiv\left(1-\beta^2\right)^{-1/2}$ is the bulk gas Lorentz
factor.  Thus, for large $\Gamma$, e.g., within the jet, $\bmath{B}$
will generally appear toroidally dominated for comoving magnetic
fields with even moderate pitch-angles.

While the black hole metric produces natural length ($GM/c^2$)
and time ($GM/c^3$) scales, the GRMHD simulations have no natural
scale for any density,
such as the plasma density, gas pressure, and magnetic field strength.
Thus we are free to renormalize the
simulation data by choosing a single arbitrary number, which we take to
be the black hole mass accretion rate, determined at the horizon by,
\begin{equation}
\dot{M} = \int_{r_h} \d\Omega \sqrt{-g}\rho u^r\,,
\end{equation}
where $g\equiv{\rm Det}(g_{\mu\nu})$ is the determinant of the metric,
$u^r$ is the contravariant radial 4-velocity and $r_h$ is the radius
of the horizon.  This gives a mass scale of $\dot{M}GM/c^3$, or
equivalently, a density scale of $\dot{M}c^3/G^2M^2$.  In the
following sections, this will be chosen to qualitatively reproduce
observations, and will vary between $10^{-5}$--$10^{-2}$ times the
Eddington rate (defined in terms of the Eddington luminosity,
$L_{\rm Edd}$,  to be $\dot{M}_{\rm Edd}\equiv10L_{\rm Edd}/c^2$).

\subsubsection{Extrapolation} \label{sec:CM:ex}
Characteristic black hole masses for the AGNs of interest range from
$10^8\Ms$ to $10^{10}\Ms$.  Given that we are seeking to address the
Faraday rotation measure distributions of AGN jets on $\pc$ scales, we are
concerned with the jet structure on scales of
$10^6GM/c^2$--$10^4GM/c^2$, respectively.  Even at the highest mass
range, the simulated region is too small.
Propagating the jet to a distance of $10^6GM/c^2$ within the GRMHD
simulation is presently computationally infeasible.
Thus we are either faced
with employing unphysically large masses ($\sim10^{11}\Ms$) or
extrapolating to larger distances.  In this paper we will do both.
Here we describe the procedure we employ for the latter, justified by
the asymptotic, power-law behavior of relativistic outflows.

Beyond the fast magnetosonic surface, highly magnetized, relativistic
jets reach the so-called monopole phase \citep{tmn09,tnm09}.  Once the
jet is in this phase the azimuthally and time-averaged fluid
quantities are very nearly self-similar and the jet moves along
conical flow lines, following power-laws for the rest-mass density,
internal energy density, components of the velocity and magnetic field, and a
logarithm for the Lorentz factor \citep{McKi:06,tmn09}.
As noted in studies of otherwise similar 2D jet simulations
that extend to larger scales ($10^4GM/c^2$) than the present 3D simulations,
the location of the fast magnetosonic surface is at $r\sim 10^2GM/c^2$ \citep{McKi:06},
suggesting that we may approximate the structure of the outflow at large
radii by advecting the simulated quantities outside of the fast
mangetosonic surface.  This is borne out by inspection of the time and
azimuthally averaged jet quantities in the full 3D simulation as
well\footnote{While the 3D GRMHD simulation does appear to follow the
  same generic structure found in the earlier large-scale 2D
  simulations reported in \citet{McKi:06}, it has insufficient radial
  dynamic range to unambiguously define the relevant asymptotic
  behaviors of the fluid quantities on its own.}.
Extrapolating the time-dependent, 3D fluid quantities
assumes that the spectrum of fluctuations is also self-similar, though
this is natural in the context of turbulence.

In axisymmetry, the self-similar structure of the fluid quantities is
constrained by a number of MHD invariants.  These are described in
detail in \citet{McKi:06}, and therefore only summarized for the
conical regime here.  As a consequence of these constraints, we must
define only one free parameter, corresponding to the rate of
dissipation, which is naturally bounded.  Generally, within the
conical regime, the fluid quantities are advected radially, and are
thus related to their values along radial flow lines.  In particular,
the radial dependencies (at fixed $\theta$) of the toroidal and
poloidal lab-frame magnetic field components are roughly given by
$B^\phi\propto\left(r\sin\theta\right)^{-1}\propto r^{-1}$
and
$B^P\propto B^\phi/r\sin\theta\propto r^{-2}$,
respectively\footnote{\citet{McKi:06} found that the jet can slightly
  de-collimate at large radii such that, e.g., $B^\phi\propto
  r^{-1.5}$ for some field lines.  However, much of the jet power
  flows along those field lines that have $B^\phi\propto r^{-1}$.
  Also, we consider a single simulation with $a/M\approx 0.92$, while
  models with much smaller $a/M$ produce a black hole jet that only
  starts to have a dominant lab-frame toroidal field beyond the~\alf
  radius (light cylinder for pulsars) located at roughly $4c
  r_h/(a/M)$, which can be quite large for small $a/M$.  However,
  below $a/M\approx 0.4$ black hole driven jets are dominated by the
  Keplerian disk-driven wind \citep{mck05}.  As a consequence, for low
  black hole spins the disk driven wind overpowers any observations of
  the black hole driven jet.  Therefore, the single $a/M\approx 0.92$
  simulation exemplifies any model that would focus on black hole
  driven jets.}.
Within the coasting regime the Lorentz factor grows with radius only
as a small power of a logarithm (i.e., roughly $(\ln(r))^{1/3}$),
and thus we approximate it as fixed, giving
$\beta^P\propto1$ along radial lines.  The toroidal velocity
is constrained via angular momentum conservation, giving
$\beta^\phi\propto\left(r\sin\theta\right)^{-1}\propto r^{-1}$ along
radial lines.  Mass conservation then implies that
$n\propto B^P/u^P \propto r^{-2}$ along radial lines.  Defining the
structure of the internal energy depends critically upon the
efficiency with which magnetic dissipation heats the baryons and electrons.
In principle, this is measured in the simulations, with \citet{McKi:06}
finding $u\propto r^{-1.3}$ along flow lines.  However, this cannot
continue indefinitely; the magnetic energy density available decreases
as $b^2/8\pi$ (where
$b^2\equiv\left|\bmath{b}\right|^2-\left(\bmath{b}\cdot\bmath{\beta}\right)^2$
is the square of the magnetic field strength in the comoving frame,
not to be confused with simply $\left|\bmath{b}\right|^2$), placing an
upper limit upon the magnetically heated baryon energy density.  A
lower limit can be found by assuming that the flow is adiabatic, giving
$u\propto n^{4/3}\propto r^{-8/3}$.  We consider both cases, adiabatic
expansion of the flow as well as setting $u\propto r^{-1.3}$ until it
reaches $\varepsilon b^2/8\pi$, for $\varepsilon\simeq0.01$--$1.0$.
In summary, within the self-similar, conical regime, the
power-law indices of $n$, $\bmath{\beta}$ and $\bmath{B}$ are
determined explicitly.  The specifics of magnetic dissipation can
effect the internal energy of the baryons, and thus we consider two
limiting cases: adiabatic and fully magnetically-thermalized flows.

In practice, to perform the extrapolation we define a piecewise-linear
remapping.  At locations within $\Rfms=10^2GM/c^2$, roughly the
location of the fast magnetosonic surface, we do no extrapolation,
returning the simulation data as is.  This ensures that we do not
extrapolate the non-asymptotic jet structure to large distances as
well as restricting the extrapolation to the outflow (since the
accretion disk is located well within this radius for the duration of
the simulation).  Outside of $\Rfms$ we linearly stretch the
coordinates so that the outer simulation boundary, $\Rsim$, is now
located at $\Rmax$.  That is, the relationship between the radius in
the physical domain, $r$, and that within the simulation, $x$, is
given by
\begin{equation}
x(r) = \left\{
\begin{aligned}
&r &&{\rm if}~r<\Rfms\\
&\Rfms + \frac{\Rsim-\Rfms}{\Rmax-\Rfms} \left(r-\Rfms\right)
&&{\rm otherwise.}
\end{aligned}
\right.
\end{equation}
In terms of this, following the discussion above, we set the values
of the fluid quantities in terms of the simulation values,
\begin{equation}
\begin{aligned}
B^\phi_{\rm ext}(r,\theta,\phi) &= \left(\frac{x}{r}\right) B^\phi(x,\theta,\phi)\\
\bmath{B}_{\rm ext}^{\bf P}(r,\theta,\phi) &= \left(\frac{x}{r}\right)^{2} \bmath{B^P}(x,\theta,\phi)\\
\beta^\phi_{\rm ext}(r,\theta,\phi) &= \left(\frac{x}{r}\right) \beta^\phi(x,\theta,\phi)\\
\bmath{\beta}_{\rm ext}^{\bf P}(r,\theta,\phi) &= \bmath{\beta^P}(x,\theta,\phi)\\
n_{\rm ext}(r,\theta,\phi) &= \left(\frac{x}{r}\right)^{2} n(x,\theta,\phi)\\
u_{\rm ext}(r,\theta,\phi) &= \left(\frac{x}{r}\right)^{-a_u} u(x,\theta,\phi)\,,
\end{aligned}
\label{eq:ext}
\end{equation}
and subsequently define $\bmath{b}$ given Equation (\ref{eq:Bb}).
We consider $a_u=-8/3$ and $-1.3$ (limiting it from above by
$\varepsilon b^2/8\pi$), corresponding to the lower and upper
limits upon the magnetic dissipation rates, respectively.  Henceforth
we will drop the subscripts upon the fluid quantities, though it should be
understood that in all cases we are referring to the extrapolated
values.

From Equation (\ref{eq:ext}) it is clear that there are important
differences between the $\pc$-scale structure of low-mass ($10^8\Ms$) and
high-mass ($10^{10}\Ms$) AGNs.  Generally, we expect that at
the same physical distance the magnetic fields of low-mass AGNs
will be even more toroidally dominated (in all locations) than their
high-mass analogs.  For similar reasons, the velocities of low-mass
AGNs will be more radially dominated than those of high-mass AGNs.

Our results are primarily dependent upon the conically expanding region
beyond $r\sim 100GM/c^2$,
except when the black hole mass is chosen to be unphysically large.
For realistic black hole masses and $\pc$-scales,
the jet magnetic field becomes toroidally dominated in both
the lab and comoving frames.  Jets that are
toroidally dominated within the comoving frame are commonly claimed
to be unstable to helical kink (screw) modes (see, e.g., the
discussion in \citealt{lpg05} regarding toroidally dominated
cylindrical jets).  However, the conical jets considered here
are stabilized by lateral expansion, resulting in causal disconnection across the jet, quenching
the growth of unstable modes \citep{mb09}.  Thus, the
substructure of conical jets at large radii
should be indicative of the interaction between the
black hole, accretion disk and jet close to the jet launching point,
which has subsequently been advected to large radii without significant additional disruption.
This is consistent with the above extrapolation procedure that essentially
advects the jet substructure to large radii.
However, we caution that such a re-scaling cannot be universally
applied to AGN jets, failing for those jets that may be significantly
affected by their environment (e.g., those associated with FRI's).  In
addition, while this implies the jet is roughly stable, some flaring
behavior due to small-scale MHD instabilities may also be expected
\citep{sbml05}.

\subsection{Rotation Measures} \label{sec:CM-RM}

\begin{figure}
\begin{center}
\includegraphics[width=\columnwidth]{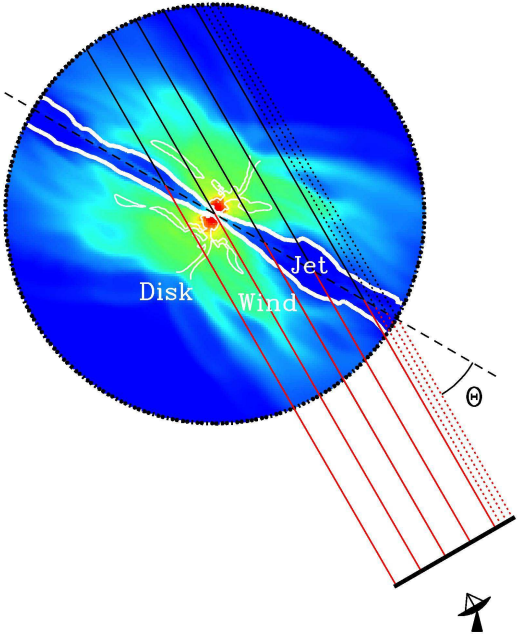}
\end{center}
\caption{Schematic representation of the relevant portions of the
  simulated jet-wind-disk system and their relationship to example
  lines of sight.  The thermal particle proper density is shown by the color
  map, extending over 9 orders of magnitude.  The disk and jet regions
  are delineated by the thin and thick white lines, respectively.  In
  between is the corona and wind (including both bound and unbound
  components).  The limited extent of the disk and wind is due both to
  finite initial extent of the disk and to rotational collimation.
  The observer is shown in the lower-right, with the image plane
  immediately above.  Lines of sight along which the $\RM$ and flux
  are integrated are shown, with the former being integrated along
  only the red portion.  Dotted lines of sight pass through the
  initial transient portion of the simulation and are thus discarded
  (see Section \ref{sec:CM-GL}).  Finally, the definition of $\Theta$
  is presented explicitly. \label{fig:schematic}}
\end{figure}

Provided the thermal electron temperature, proper density, four-velocity and
magnetic field strength and geometry, we can now compute the $\RM$
along any given line of sight.  Following \citet{Brod-Loeb:09b}, we do
this directly via
\begin{equation}
\RM = 0.812\times10^6 \int \frac{f(T) n}{g^2}
\left(\bmath{\hat{k}}-\bmath{\beta}\right)\cdot\bmath{b}d\ell
\,\rad\,\m^{-2}\,,
\label{eq:RMdef}
\end{equation}
where
$g\equiv\Gamma\left(1-\bmath{\beta}\cdot\bmath{\hat{k}}\right)$
and we have supplemented this with a finite temperature
term\footnote{The rate of Faraday rotation depends upon the nature of
  the underlying electromagnetic eigenmodes of the plasma.  For cold
  plasmas these are circular, reproducing the standard result.  For
  ultra-relativistic plasmas these are substantially elliptical,
  strongly suppressing any Faraday rotation.}, given by,
\begin{equation}
f(T) =
\frac{1}{\gamma^2}
+
\frac{\gamma-1}{2\gamma^3}
\log\left(\gamma\right)
\quad\text{where}\quad
\gamma = \max\left(1,\frac{kT}{m_ec^2}\right)\,,
\end{equation}
where $\gamma$ is the thermal particle Lorentz factor, not to be
confused with the bulk Lorentz factor of the out-flowing gas
\citep{Huan_etal:08,Shch:08}.  Note that the $\RM$, as we have defined
it, depends solely upon quantities that have been simulated, with the
integration geometry shown in Figure \ref{fig:schematic}.  All
integrals are performed in flat space, ignoring any contributions from
gravitational lensing, an approximation that is well-justified at the
distances responsible for the resolved polarized emission.  With this
restriction, Equation (\ref{eq:RMdef}) is correct for relativistically
moving Faraday screens.

\begin{figure}
\begin{center}
\includegraphics[width=\columnwidth]{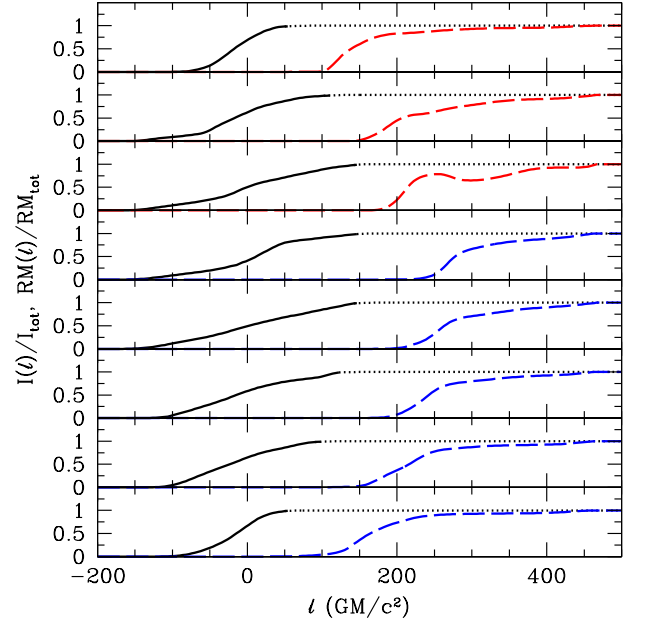}
\end{center}
\caption{Fractional integrated intensity (black) and rotation measure
  (red \& blue dashed lines, corresponding to positive and negative
  $\RM$s) as a function of distance as measured from the jet
  mid-plane along a number of exemplar lines of sight, which were
  chosen such that they span the width of the jet.  After reaching
  99\% of its final value, the intensity is shown by a dotted line.
  In all cases, along any given line of sight the region responsible
  for the $\RM$ is well-separated from that responsible for producing
  the observed emission. \label{fig:IRM}}
\end{figure}

\begin{figure*}
\begin{center}
\includegraphics[width=\textwidth]{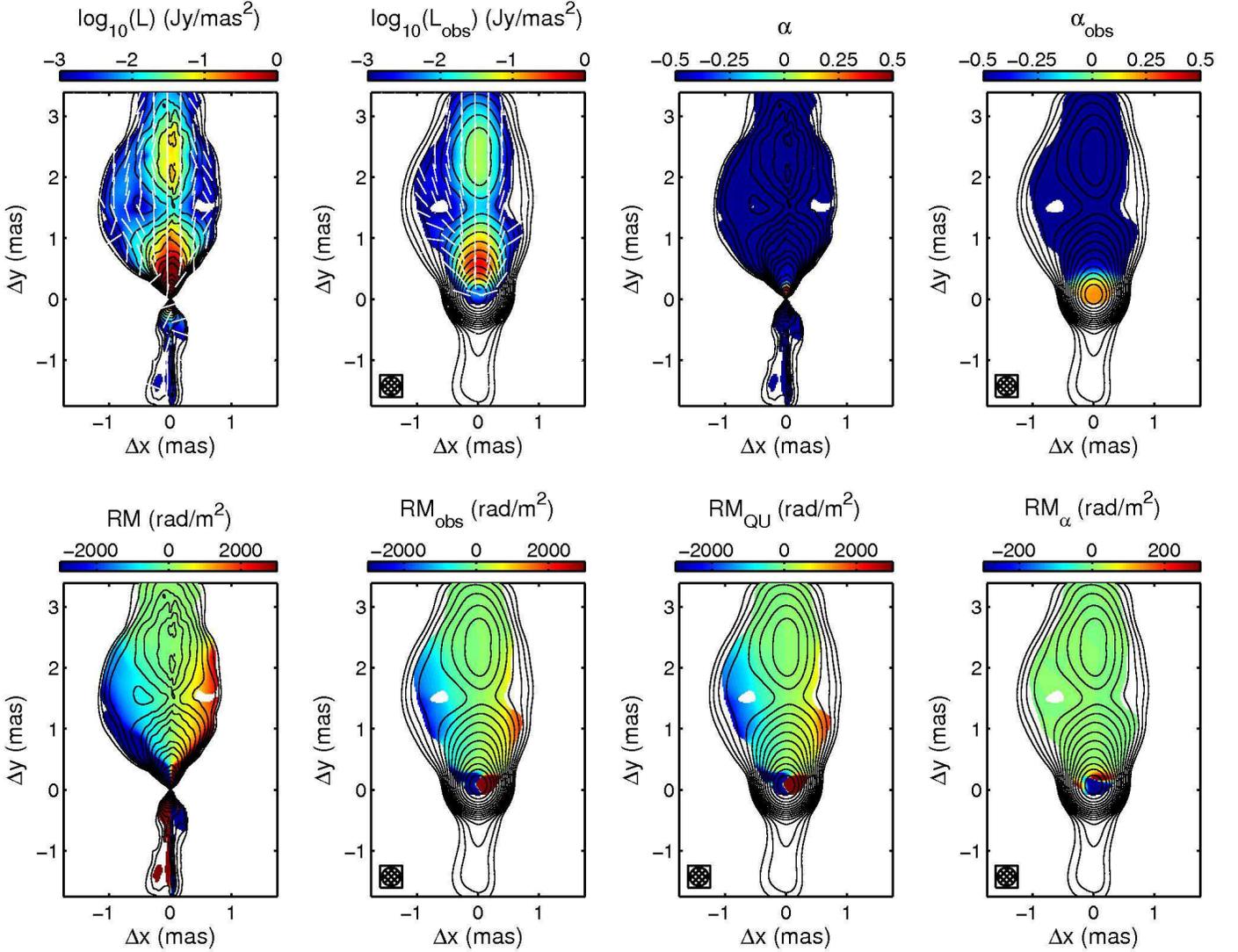}
\end{center}
\caption{Ideal-resolution and beam-convolved polarized flux with
  unrotated polarization ticks (upper-left), spectral index
  (upper-right) and $\RM$ maps (lower-left) for our canonical jet
  model as observed at $15\,\GHz$ (see Table \ref{tab:simparams} for a
  complete list of relevant parameters).  In addition the
  contributions from the two terms in Equation (\ref{eq:RMavg}) are
  shown explicitly (lower-right).  In all panels total flux contours
  are overlaid, in logarithmic factors of 2, with the minimum contour
  corresponding to $0.98\,\mJy \mas^{-2}$, and all other quantities are
  clipped at a polarized flux of $1\,\mJy\,\mas^{-2}$.  Beam-convolved
  maps show the beam size ($0.3\,\mas$ FWHM) in their lower-left
  corner.
\label{fig:exIRMmaps}}
\end{figure*}

The integral in Equation (\ref{eq:RMdef}) is performed from a vertical
mid-plane, to a distant image plane, orthogonal to the line of sight.
In principle, this overestimates the observed $\RM$ because it ignores
depolarization along the line of sight resulting from Faraday rotation within
the emitting region.  However, in practice the emitting and rotating
regions are well-separated, as illustrated explicitly in Figure
\ref{fig:IRM}, which shows the fractional local contributions to the
Faraday rotation and emission (see the following section) along a
representative line of sight.  That this is the case is also supported
directly by the observation of significant jet polarization fractions,
ruling out substantial Faraday rotation within the emitting region
\citep{burn66,gardner66}.  To generate an $\RM$ map, we then repeat
this process many times at different positions on a rectilinear grid
located on the image plane. Examples of the resulting maps can be
found in Figure \ref{fig:exIRMmaps} and Section \ref{sec:Implications}.

\subsection{Jet Synchrotron Emission} \label{sec:CM-JSE}
While we may now compute the $\RM$ along any line of sight, given a
complete description of the circum-jet material, in practice $\RM$s
can be measured only for those lines of sight along which the jet is
sufficiently bright.  As a consequence, in order to compare with
observations of AGN jets, we must also model the jet emission.  This
will serve two purposes: first as a mask upon the $\RM$-maps,
identifying the regions that are observable, and second as a weight
for determining the beam-convolved $\RM$ maps, as described in Section
\ref{sec:CM-BC}.

\subsubsection{Particle Acceleration}
Arising from nonthermal synchrotron emission within the jet itself,
the emission is much more poorly constrained by the simulations.  The
difficult arises primarily from the lack of an ab initio
understanding of the mechanism by which the radiating nonthermal
electrons are accelerated.  Here we consider a simple, qualitative
model, appropriate when the accelerated electrons are the result of
the dissipation of MHD turbulence and/or magnetic reconnection within the highly-magnetized
jet\footnote{Internal shocks have also been suggested as a mechanism
  for producing the observed nonthermal particles.  We have
  considered a variety of shock-acceleration schemes, based upon
  identifying the locations of resolved shocks within the simulation
  and setting the nonthermal energy density to a shock-strength
  dependent fraction of the post-shock internal energy.  These
  generally result in disk-dominated emission, and are incapable of
  reproducing the observed extended jet emission.  This implies that
  if shocks are responsible for producing a substantial portion of the
  nonthermal electrons, they must be microscopic or produced
  by an interaction between the jet and a surrounding ambient medium
  that is not included in the simulation we consider.}.
In this case the
reservoir of energy that is available to be tapped via particle
acceleration is that associated with the electromagnetic field.  Thus, we
set the comoving energy density in nonthermal electrons, $\un$,
proportional to that of the simulation electromagnetic field, $b^2/8\pi$.
This assumption is consistent with minimum energy arguments
used to infer the expected density of emitting particles
and magnetic field strengths for emission due to synchrotron emission
within AGNs \citep{burbidge56}.
We further restrict the particle acceleration by requiring that it
occur only in magnetically dominated regions, i.e., where
$b^2/8\pi>\rho c^2$, which ensures that particles can be efficiently
accelerated to large Lorentz factors via MHD turbulence and/or
magnetic reconnection \citep{lyubarsky05,lyutikov06}.
That is, explicitly, we set
\begin{equation}\label{unfull}
\un = \eta \frac{b^2}{4\pi} \left[1+\exp\left(\frac{\rho c^2}{b^2/8\pi}\right)\right]^{-1}\,,
\end{equation}
and $\eta$ is the efficiency with which magnetic energy is converted into
that in nonthermal particles [recall that
$b^2\equiv\left|\bmath{b}\right|^2-\left(\bmath{b}\cdot\bmath{\beta}\right)^2$].
In magnetically dominated regions this
reduces to $\un = \eta b^2/8\pi$, and is exponentially suppressed
otherwise.  This is motivated first by the physical assumption that
conversion of electromagnetic energy into nonthermal particles is less
efficient when the electromagnetic energy per particle is low
\citep[e.g., relativistic magnetic reconnection requires, roughly,
$b^2/(8\pi)\gg \rho c^2$,][]{lu03}.  However, in practice
relaxing this suppression results is broad flux distributions that are
inconsistent with observations of AGN jets.

As a physically motivated alternative, one might choose $\un$ to be
proportional to the square of the comoving current density,
indicative of where current sheets are dissipating \citep{lpg05}.
However, the current density in the simulations is unlikely to be
properly resolved and outside the current sheet the properties of the
current layer are determined by the comoving electromagnetic energy
density, suggesting that $\un\propto b^2$ also roughly tracks the
importance of current sheets if they were present.  We leave the
consideration of other prescriptions for the nonthermal particle
density for future work.

We further assume a power-law electron distribution, with
index $-2$ (resulting in a spectral index of 0.5) and minimum particle
Lorentz factor, $\gamma_{\rm min}$, of $10^2$.  Since our
primary concern is to identify bright and dim regions, moderate
changes in either of these do not significantly change our results.

\subsubsection{Polarized Radiative Transfer in Bulk Flows} \label{sec:CM:PRTiBF}
Due to the strong magnetic field within the jet and the relatively
low density, the jet emission will be dominated by synchrotron.
On $\pc$ scales, AGN jets are generally optically thin, while the bright
radio core is optically thick. Thus, synchrotron self-absorption is
critical to obtaining total jet fluxes and modelling the jet spectral
index distribution.  Furthermore, in order to compute the
beam-convolved $\RM$s we will need intrinsic Stokes $I$, $Q_0$ and
$U_0$.  Thus we are necessarily faced with computing the
self-absorbed, polarized emission.

The transfer of polarized radiation through relativistic environments
has been discussed by a variety of authors and thus here
we only summarize the procedure
\citep[see, e.g.,][]{Lind:66,Conn-Star:77,Conn-Star:80,Laor-Netz-Pira:90,Jaro-Kurp:97,Brom-Meli-Liu:01,Brod-Blan:04,Beck-Done:05,Brod:06,Brod-Loeb:06b}.  This is simplified somewhat by the
assumption that within the regions of interest we may approximate the
spacetime as being flat, requiring only that we consider the
relativistic motion of the emission region.  This results in the Doppler
boosting and beaming of the emission as well as relativistic aberration
of the polarization.  All of these can be taken into account by noting
that the photon occupancy numbers, $\propto (I,Q_0,U_0)/\nu^3$,
are Lorentz scalars, allowing us to rewrite the standard radiative
transfer equation in the emitting frame.  From this we obtain,
\begin{equation}
\frac{d}{d\ell}
\left(\begin{aligned}
I\\ Q_0\\ U_0
\end{aligned}\right)
=
\bar{j}_\nu
\left(\begin{aligned}
1\\ \epsilon_Q\\ \epsilon_U
\end{aligned}\right)
-
\bar{\alpha}_\nu
\left(\begin{aligned}
1 && \zeta_Q && \zeta_U\\
\zeta_Q && 1 && 0\\
\zeta_U && 0 && 1\\
\end{aligned}\right)
\left(\begin{aligned}
I\\ Q\\ U
\end{aligned}\right)\,,
\label{eq:rt}
\end{equation}
where $\bar{j}_\nu\equiv j_{g\nu}/g^2$ and
$\bar{\alpha}_\nu\equiv g\alpha_{g\nu}$ are Lorentz covariant analogs
of the lab-frame emission and absorption coefficients ($j_\nu$ and
$\alpha_\nu$, respectively), which with $\epsilon_{Q,U}$ and
$\zeta_{Q,U}$, are discussed in more detail below.

Generally, the form of $\bar{j}_\nu$ depends upon
the locations of the minimum and maximum nonthermal particle Lorentz
factors (we have specified only the former), and the location of a
cooling break (located where particle acceleration balances
synchrotron cooling).  However, the cooling time at radio wavelengths
exceeds the jet propagation time at $\pc$ scales, and thus we may
safely ignore cooling in the computation of the $\pc$-scale radio
images.  Furthermore, while the maximum particle Lorentz factor
generally depends upon the details of the particle acceleration
mechanism, typical estimates, given the jet magnetic field strength,
are much larger than the Lorentz factors of the particles responsible
for the radio emission.  This leaves the spectral break associated
with $\gamma_{\rm min}$, giving
\begin{multline}
\bar{j}_\nu
=
C_j\frac{\un}{g^2\gamma_{\rm min}} \nu_B \sin^{1.5}\vartheta\\
\times
\left\{
\begin{aligned}
&\left(\frac{\nu_B}{g\nu_{\min}}\right)^{0.5}\left(\frac{\nu}{\nu_{\min}}\right)^{1/3}
&&
\text{if~} \nu<\nu_{\min}\\
&\left( \frac{\nu_B}{g\nu} \right)^{0.5}
&&
\text{otherwise,}\\
\end{aligned}
\right.
\label{eq:jnu}
\end{multline}
in which
$C_j\simeq60\,\Jy\,\cm^2\,\erg^{-1}\,\Hz^{-1}\,\Sr^{-1}$,
$\nu_B\equiv eb/2\pi m_e c$ is the cyclotron frequency in the comoving
frame,
$\sin\vartheta \equiv \sqrt{ g^2b^2 - \left[\left(\bmath{\hat{k}}-\bmath{\beta}\right)\cdot\bmath{b}\right]^2 }\Big/gb$
is the sine of the angle between the line of sight and the magnetic
field in the comoving frame,
and $\nu_{\min}\equiv\gamma_{\rm min}^2\nu_B\sin\vartheta/g$ is the
frequency associated with spectral break due to $\gamma_{\rm min}$.
Given $\bar{j}_\nu$, we may determine $\bar{\alpha}_\nu$ via
Kirchhoff's law, giving,
\begin{multline}
\bar{\alpha}_\nu
=
C_\alpha g\frac{\un}{\gamma_{\rm min}}\frac{\sin^2\vartheta}{\nu_B}\\
\times
\left\{
\begin{aligned}
&\left(\frac{\nu_B}{g\nu_{\min}}\right)^{3}\left(\frac{\nu}{\nu_{\min}}\right)^{1/3}
&&
\text{if~} \nu<\nu_{\min}\\
&\left( \frac{\nu_B}{g\nu} \right)^{3}
&&
\text{otherwise,}\\
\end{aligned}
\right.
\label{eq:alphanu}
\end{multline}
where $C_\alpha\simeq1.3\times10^6\,\cm^{-1}$.

The polarization terms are affected by both the relativistic
aberration associated with the bulk motion of the jet and the
orientation of the magnetic field.  Both of these can be addressed by
parallely propagating an orthonormal tetrad along the ray trajectory,
defined at the image plane such that two of the basis vectors are
aligned with the directions defining the observed Stokes parameters
(and thus orthogonal to the line of sight).  In principle, this basis
may then be boosted into the emitting frame, directly accounting for
relativistic aberration, and then compared with the projected
direction of the magnetic field.  In practice, it is sufficient to
define the angle, $\xi$, between the initially vertical basis vector
(associated with $Q$) and the projected magnetic field direction, in
the emitting frame.  As shown in Appendix \ref{app:RA}, this is defined up to a
sign by
\begin{equation}
\cos\xi =
\frac{
\bmath{\hat{z}}
\cdot
\left(\bmath{1}-\bmath{\hat{k}}\bmath{\hat{k}}\right)
\cdot
\left(\bmath{b}+\Gamma\bmath{\beta}b\cos\vartheta\right)
}{
\sqrt{1-\left(\bmath{\hat{z}}\cdot\bmath{\hat{k}}\right)^2}b\sin\vartheta
}\,.
\label{eq:polabb}
\end{equation}
The sign of $\xi$ may then be obtained by a similar construction
obtaining $\sin\xi$ from the relationship between the initially
horizontal basis vector and the projected magnetic field in the
emitting frame.  However, it is sufficient to note the direction of
$\bmath{b}$ in the lab frame since boosting preserves the ordering of
vectors, e.g., if $\bmath{b}$ is to the right of $\bmath{\hat{z}}$ (as
seen by looking down the line of sight) in the emitting frame,
then it will be so in all frames.  Thus,
$\sin\xi = \sgn\left[\bmath{\hat{b}}\cdot\left(\bmath{\hat{z}}\times\bmath{\hat{k}}\right)\right]\sqrt{1-\cos^2\xi}$.
In terms of these, the polarization terms are given by
\begin{equation}
\begin{gathered}
\epsilon_Q = -\frac{9}{13}\cos2\xi\,,\quad
\epsilon_U = -\frac{9}{13}\sin2\xi\,,\\
\zeta_Q = -\frac{3}{4}\cos2\xi\,,\quad
\zeta_U = -\frac{3}{4}\sin2\xi\,,\\
\end{gathered}
\label{eq:polterms}
\end{equation}
where the particular numerical coefficients are set by the choice of
the electron spectrum index and the overall sign reflects the fact
that in the emitting frame the polarization is orthogonal to the
projected magnetic field.

As with the expression for the $\RM$, Equations
(\ref{eq:rt}--\ref{eq:polterms}) are Lorentz covariant.  Equation (\ref{eq:rt})
is integrated throughout the emitting region along a given line of
sight ending upon the image plane.  However in practice, the emission
is limited to the jet and associated accretion disk.  Again, this
procedure is repeated at a number of points upon the image plane,
producing an image of the jet itself.  Upon choosing a black hole
mass, $\dot{M}$, distance and beam size, we generate a normalized
radio map of the polarized jet emission, examples of which may be
found in Figure \ref{fig:exIRMmaps} and Section \ref{sec:Implications}.

In the following section we will find that the spatial distribution of
the spectral index of the polarized emission, $\alpha_L$, will be
important in determining the observed $\RM$ maps.  As a consequence,
we also construct $\alpha_L\equiv-d\log\sqrt{Q_0^2+U_0^2}/d\log\nu$
and $\alpha\equiv-d\log I/\d\log\nu$ by
differencing polarized emission maps at nearby frequencies.  Both
$\alpha_L$ and $\alpha$ (an example of which is shown in Figure
\ref{fig:exIRMmaps}) are generally core dominated with a size
indicative of the accretion rate.

\subsection{Beam Convolution} \label{sec:CM-BC}
Radio maps of AGN jets are generally under-resolved, with the
consequence that a great deal of the jet substructure is smoothed out
by the radio beam.  However, in the case of the $\RM$ maps, we must
take some care since $\RM$ is not observed directly.  In this section
we describe how we perform an appropriate beam convolution operation
on the $\RM$ maps.  We begin with a general discussion regarding
finite beamwidth $\RM$ maps, and then move onto the particulars of the
scheme we utilize in this paper.

\subsubsection{Finite Beamwidth $\RM$ Maps in General}
It is possible to infer the beam-convolved $\RM$ maps without
producing fully simulated observations, i.e., without explicitly
producing polarized images of the jet at multiple wavelengths.  This
is done by careful considering what is, and is not, directly
measured.  In practice, the $\RM$ is a derived quantity, computed from
directly measured Stokes parameters $Q$ and $U$.  Along with $I$, these
are limited by the effective telescope beam, with the observed map
being related to the infinite-resolution values via a convolution:
\begin{equation}
\left<I\right>(\bmath{\Omega})
=
\int d\Omega\, B(\bmath{\Omega}'-\bmath{\Omega}) I(\bmath{\Omega}')\,,
\end{equation}
with similar relations for $Q$, $U$ and $V$, in which
$B(\bmath{\Omega})$ is the beam profile (which we will ultimately take
to be Gaussian).  In contrast, since $\RM$ is non-linearly related to
$Q$ and $U$, the relationship between the finite-beam and the
ideal-resolution $\RM$s is not a simple convolution.  Nevertheless,
as we show below, it may be written as the combination of weighted
convolutions.

The effect of an intervening Faraday screen is to rotate $Q$ into $U$
and vice versa:
\begin{equation}
\begin{aligned}
Q &= L \cos\left(2\Psi_0 + 2\RM\lambda^2\right)\\
U &= L \sin\left(2\Psi_0 + 2\RM\lambda^2\right)\,,
\end{aligned}
\label{eq:QUdef}
\end{equation}
where $L=\sqrt{Q_0^2+U_0^2}$ is the intrinsic (ideal-resolution)
polarized flux, $Q_0=L\cos\left(2\Psi_0\right)$ and
$U_0=L\sin\left(2\Psi_0\right)$.  The observed $Q$ and $U$ are then
\begin{equation}
\begin{aligned}
\left<Q\right> &= \left<L \cos\left(2\Psi_0 + 2\RM\lambda^2\right)\right>\\
\left<U\right> &= \left<L \sin\left(2\Psi_0 + 2\RM\lambda^2\right)\right>\,,
\end{aligned}
\end{equation}
and is now explicitly dependent upon the spatial distribution of the $\RM$
and $\lambda$.  The observed polarization angle is
\begin{equation}
\Psi_{\rm obs} = \frac{1}{2} \arctan\left(\frac{\left<U\right>}{\left<Q\right>}\right)\,.
\end{equation}
This implies that the measured $\RM$ is then
\begin{equation}
\begin{aligned}
\RMo
&\equiv
\frac{d\Psi_{\rm obs}}{d\lambda^2}\\
&=
\frac{1}{2\left(\left<Q\right>^2 + \left<U\right>^2\right)}
\left(
\left<Q\right>
\frac{d\left<U\right>}{d\lambda^2}
-
\left<U\right>
\frac{d\left<Q\right>}{d\lambda^2}
\right)\,.
\end{aligned}
\end{equation}

Ignoring the wavelength dependencies in the beam size (which are
explicitly accounted for with match-resolution images, produced by
convolving high-frequency images with the lowest frequency-interval
beam) $d/d\lambda^2$ commutes with the beam convolution.
Thus, with Equations (\ref{eq:QUdef}),
\begin{equation}
\frac{d\left<Q\right>}{d\lambda^2}
=
\left<\frac{dQ}{d\lambda^2}\right>
=
\left<Q \frac{\alpha_L}{2\lambda^2}\right>
-
2 \left<U\RM\right>
\,,
\end{equation}
where we have assumed that
$\Psi_0$ is wave-length independent (i.e., Faraday rotation is the
{\em only} frequency dependence in $\Psi$).  Similarly,
\begin{equation}
\frac{d\left<U\right>}{d\lambda^2}
=
\left<\frac{dU}{d\lambda^2}\right>
=
\left<U \frac{\alpha_L}{2\lambda^2}\right>
+
2 \left<Q\RM\right>
\,.
\end{equation}
As a consequence,
\begin{equation}
\begin{aligned}
\RMo
&=
\frac{
\left<Q\right>\left<Q\RM\right> + \left<U\right>\left<U\RM\right>
}{
\left<Q\right>^2 + \left<U\right>^2
}\\
&\qquad+
\frac{
\left<Q\right>\left<U\alpha_L\right> - \left<U\right>\left<Q\alpha_L\right>
}{
4\lambda^2\left(\left<Q\right>^2 + \left<U\right>^2\right)
}\,.
\label{eq:RMavg}
\end{aligned}
\end{equation}
The first term corresponds to a weighted average of the $\RM$ across
the beam.  Unresolved correlations between the polarized flux and the
$\RM$ can produce substantial discrepancies between the intrinsic
and observed $\RM$s.
At this point we can identify some of the important differences
between observations of $\RM$s and the polarized fluxes.  If the $\RM$
is fixed across the beam we have $\RMo=\RM$ trivially.  This is true
even if there are rapid variations in the intrinsic
polarization\footnote{Though in this case the measured polarized flux
  may be so small as to make measuring the $\RMo$ no longer
  feasible.}.  The converse is not true; order $2\pi/\lambda^2$
variations in $\RM$ across the beam results in the well-known Faraday
beam depolarization.  More important are correlated variations in the
polarized intensity and the $\RM$.  Since $\RMo$ is a weighted
average, clear large-scale structures in the $\RM$ can be obscured by
the presence of beam-scale structure in $L$.  Thus even a uniform
$\RM$ gradient can appear non-uniform at the beam-scale when
illuminated by a spatially variable source, and in particular near the
radio core where all quantities are changing on scales small in
comparison to the observing beam.

The second term is associated with spatial variations in the
polarized spectral index, vanishing identically when $\alpha_L$ is
constant.  This is due to the potential for frequency dependent
competition among unresolved regions with different intrinsic
polarizations.
Spatially varying spectral indices have been found in a number of
sources.  Most common are swings from $\alpha_L\simeq1$--$0.5$ within
the optically thick AGN core and settling to $\alpha_L\simeq-0.5$ at
angular distances of more than $\sim1\,\mas$, where the jet is
optically thin, though cases where similar magnitude variations across
the jet occur are known, presumably due to interactions with the AGN
environment \citep{OSul-Gabu:09a,OSul-Gabu:09b}.  For comparison with
these we also compute the beam-convolved spectral index:
\begin{equation}
\begin{aligned}
\alpha_{\rm obs}
&\equiv
\frac{d\log \langle I\rangle}{d\log\lambda}
=
\frac{\langle I \alpha \rangle}{\langle I \rangle}\,.
\end{aligned}
\end{equation}
Examples may be found in Figure \ref{fig:exIRMmaps} and Section
\ref{sec:Implications}.

\subsubsection{Beam-Convolved $\RM$ Gradients}
We will remark upon the implications of finite beams for the
$\RM$-maps of AGN jets in detail in the following sections.  However,
a simple, qualitative example directly illustrates the importance of
beam convolution upon the search for, and study of,
transverse $\RM$ gradients within AGN jets.  We will find that generally
it is crucial to resolve the transverse structure in both the emission
and the $\RM$ distribution in order to obtain accurate measurements of
the true $\RM$s.

We briefly consider a uniformly polarized 1D, cylindrically symmetric
jet with a Gaussian transverse flux profile with width $\sigma_J$, a
circular Gaussian beam with width $\sigma_B$, and a linear
$\RM$-profile.
The typical magnitude of the $\RM$s under consideration is
$10^3\,\rad\,\m^{-2}$, which at $15\,\GHz$ produces Faraday rotations of
$\sim0.4\,\rad$.  For the present purpose this is sufficiently small
that it may be safely neglected (though we have confirmed this
via explicit computation). Thus, without loss of generality we may set,
\begin{equation}
\RM = \RM_0' x\,,\quad
B = \frac{\e^{-x^2/2\sigma_B^2}}{\sqrt{2\pi}\sigma_B}\,,\quad
Q = \frac{\e^{-x^2/2\sigma_J^2}}{\sqrt{2\pi}\sigma_J}\,,
\end{equation}
where we have rotated the axes defining the Stokes parameters to force
$U$ to vanish.  With these the various beam-convolved quantities may
be computed analytically, giving
\begin{equation}
\begin{aligned}
\langle Q\rangle(x)
&=
\int_{-\infty}^\infty dy B(y-x) Q(y)
=
\frac{\e^{-x^2/2(\sigma_J^2+\sigma_B^2)}}{\sqrt{2\pi(\sigma_J^2+\sigma_B^2)}}\\
\langle Q\RM\rangle(x)
&=
\int_{-\infty}^\infty dy B(y-x) Q(y) \RM(y)
=
\frac{\RM_0'x\sigma_J^2}{\sigma_J^2+\sigma_B^2}\langle Q\rangle(x)\,,
\end{aligned}
\end{equation}
which then implies
\begin{equation}
\RMo = \RM \frac{\sigma_J^2}{\sigma_J^2+\sigma_B^2}\,.
\label{eq:RMgrad}
\end{equation}
That is, for linear $\RM$ gradients, the beam-convolved $\RM$s are
generally smaller than the true values, and is a relatively strong
function of the ratio of jet width to beam size.  As a
consequence, when the transverse jet structure is not resolved, even
apparently linear $\RM$ gradients will not be accurate measures of the
true jet $\RM$s.  We will find that Equation (\ref{eq:RMgrad}) will
be useful in interpreting the significance of unresolved, but
nevertheless robust, $\RM$ gradients.

\subsubsection{Computing Beam-Convolved $\RM$ Maps in Practice}
Where the gradients in $\RM$ are sufficiently high, the
procedure by which we have computed the $\RM$, $(I,Q_0,U_0)$ and $\alpha_L$
maps fails to resolve the rapidly varying polarizations.  That is,
when the image grid scales (generally much smaller than the beam size)
are $\Delta\Omega_x$ and $\Delta\Omega_y$, if
$\Delta\Omega_x d\RM/d\Omega_x\gtrsim\pi$ or
$\Delta\Omega_y d\RM/d\Omega_y\gtrsim\pi$, we
must supplement the inferred $Q$ and $U$ maps with a sub-grid model.
Since the grid scale is much smaller than the typical beam scale, we
may assume that $B(\bmath{\Omega})$ is roughly constant over a single
pixel, and thus the primary effect corresponds to a grid-scale analog
of beam depolarization.  In this case, it is straightforward to show
that the effective polarization fraction is
\begin{equation}
m_L
=
m_{L,0}\,
\sinc\left(\Delta\Omega_x\frac{d\RM}{d\Omega_x}\lambda^2\right)
\sinc\left(\Delta\Omega_y\frac{d\RM}{d\Omega_y}\lambda^2\right)\,.
\label{eq:mLdef}
\end{equation}
Thus, the procedure for producing the a finite-beam $\RM$ map is to
first compute the ideal resolution $\RM$, $(I,Q_0,U_0)$ and $\alpha_L$
maps, pixel-depolarize these maps using Equation (\ref{eq:mLdef}), and
perform the convolutions in Equation (\ref{eq:RMavg}) assuming a
Gaussian beam.
Examples of the resulting $\RM$ maps are presented in Figure
\ref{fig:exIRMmaps} and Section \ref{sec:Implications}.

\subsection{Geometric Limitations} \label{sec:CM-GL}
There are necessarily geometric limitations upon the jet inclinations,
$\Theta$, for which we can produce $\RM$ maps.  For $\Theta$ less than the
jet half-opening angle, $\sim 3^\circ$ at $\Rmax$, we are
looking down the jet directly.  As a consequence, the resulting
$\RM$s are indicative of initial transients in the simulation, and of
the radio lobe in real systems.  Thus, we restrict ourselves to
$\Theta\ge10^\circ$.

However, these transients combined with the finite simulation time
constrain the angular size of the image region more generally.  The
projected length of the jet is roughly
$L_{\rm proj}\simeq\Rmax\sin\Theta$.  Lines of sight which pass
through regions more distant than this, shown by the dotted lines in
Figure \ref{fig:schematic}, are necessarily indicative of the initial
transient period.  Thus, we limit ourselves to lines of sight that
pass through the non-transient portions of the jet, and therefore
restrict the region on the image plane we ultimately consider.  Since
this constraint becomes more severe at small $\Theta$, becoming
impossible to satisfy when $\Theta$ is equal to the half-opening angle
of the jet, the credible portions of the $\RM$ maps are necessarily
different sizes for different inclinations.

\section{Jet $\RM$ Maps and their Implications} \label{sec:Implications}

\begin{deluxetable*}{ccccccccccc}\tabletypesize{\small}
\tablecaption{Jet model parameters\label{tab:simparams}}
\tablehead{
\colhead{Figure} &
\colhead{$\nu$} &
\colhead{beam\tablenotemark{a}} &
\colhead{$\Theta$} &
\colhead{$\Rmax$} &
\colhead{$M$} &
\colhead{$\dot{M}$} &
\colhead{$\eta$} &
\colhead{$a_u$} &
\colhead{$\varepsilon$} &
\colhead{$F_{15\,\GHz}$}\\
 & ($\GHz$) & ($\mas$) & & ($\Rsim$) & ($10^8\Ms$) &
($\dot{M}_{\rm Edd}$) & & & & ($\Jy$)
}
\startdata
\ref{fig:exIRMmaps},\ref{fig:azi},\ref{fig:mass},\ref{fig:diss},\ref{fig:struc},\ref{fig:sRM2} & 15 &  0.3 & $20^\circ$ & $10^3$ &    3 &      0.01 & 0.02 & $-8/3$ &   -- & 3.83\\
\hline
\ref{fig:inc},\ref{fig:sRM2} & 15 &  0.3 & $10^\circ$ & $10^3$ &    3 &     0.005 & 0.02 & $-8/3$ &   -- & 3.77\\
\ref{fig:inc},\ref{fig:sRM2} & 15 &  0.3 & $30^\circ$ & $10^3$ &    3 &     0.018 & 0.02 & $-8/3$ &   -- & 3.89\\
\hline
\ref{fig:mass} & 15 &  0.3 & $20^\circ$ & $10^2$ &   30 & $10^{-3}$ & 0.02 & $-8/3$ &   -- & 4.00\\
\ref{fig:mass},\ref{fig:sRM2} & 15 &  0.3 & $20^\circ$ & $10$   &  300 & $10^{-4}$ & 0.02 & $-8/3$ &   -- & 6.67\\
\ref{fig:mass} & 15 &  0.3 & $20^\circ$ & $1$    & 3000 & $10^{-5}$ & 0.02 & $-8/3$ &   -- & 12.7\\
\hline
\ref{fig:diss},\ref{fig:sRM2} & 15 &  0.3 & $20^\circ$ & $10^3$ &    3 &      0.01 & 0.02 & $-1.3$ &    1 & 3.83\\
\ref{fig:diss} & 15 &  0.3 & $20^\circ$ & $10^3$ &    3 &      0.01 & 0.02 & $-1.3$ &  0.1 & 3.83\\
\ref{fig:diss} & 15 &  0.3 & $20^\circ$ & $10^3$ &    3 &      0.01 & 0.02 & $-1.3$ & 0.01 & 3.83\\
\hline
\ref{fig:freq} & 5  &  0.3 & $20^\circ$ & $10^3$ &    3 &      0.01 & 0.02 & $-8/3$ &   -- & 3.35\\
\ref{fig:freq} & 8  &  0.3 & $20^\circ$ & $10^3$ &    3 &      0.01 & 0.02 & $-8/3$ &   -- & 3.57\\
\ref{fig:freq} & 22 &  0.3 & $20^\circ$ & $10^3$ &    3 &      0.01 & 0.02 & $-8/3$ &   -- & 3.97\\
\ref{fig:freq},\ref{fig:sRM2} & 43 &  0.3 & $20^\circ$ & $10^3$ &    3 &      0.01 & 0.02 & $-8/3$ &   -- & 4.15\\
\hline
\ref{fig:beam} & 15 & 0.04 & $20^\circ$ & $10^3$ &    3 &      0.01 & 0.02 & $-8/3$ &   -- & 3.83\\
\ref{fig:beam} & 15 &  0.1 & $20^\circ$ & $10^3$ &    3 &      0.01 & 0.02 & $-8/3$ &   -- & 3.83\\
\ref{fig:beam},\ref{fig:sRM2} & 15 &  0.9 & $20^\circ$ & $10^3$ &    3 &      0.01 & 0.02 & $-8/3$ &   -- & 3.83
\enddata
\tablenotetext{a}{FWHM}
\end{deluxetable*}

In this section we present $\RM$ maps for a variety of model,
observational and theoretical parameters.  These include variations in
AGN model parameters (e.g., the jet orientation, black hole mass and
magnetic dissipation rate; \S\ref{sec:JRMM-AP}), observing parameters
(observation frequency and beam size; \S\ref{sec:JRMM-OP}) and the
structure of the Faraday screen (\S\ref{sec:JRMM-JS}).  To ensure that
we are comparing physically similar systems, we define a canonical
model: A central black hole with mass $3\times10^8\Ms$ accreting at
$1\%$ of the Eddington rate located at a
distance of $300\,\Mpc$ and viewed at an inclination of
$\Theta=20^\circ$.  For the extrapolation we choose
$\Rmax=10^3\Rsim=10^6GM/c^2\simeq15\,\pc$, and select $a_u=-8/3$,
corresponding to the adiabatic advection model.  Finally we choose
$\eta=0.02$ such that the jet flux is roughly $4\,\Jy$ at $15\,\GHz$.
These choices are not unique, and we will consider explicit departures
from this canonical model in all respects; a list of the parameters
of the models we present are listed in Table \ref{tab:simparams}.

Figure \ref{fig:exIRMmaps} illustrates the general morphologies of the
flux, polarization, spectral index and $\RM$ associated with our
simulated jets.  As in all subsequent $\RM$ maps, we show the
polarization properties only where they are likely to be observed,
requiring the polarized flux exceed $1\,\mJy\,\mas^{-2}$~\footnote{We choose a
limit on a fixed surface brightness, as opposed to $\Jy$ per beam, so that
we may later compare maps with different beam sizes on equal footing.}.
  Specifically, in all cases we see an optically thick,
small core that dominates the total and polarized emission (see the
left-most and right-center upper panels).  The ordered magnetic field
within the jet results in an ordered polarization orientation,
departing from the preferred direction only near the jet edge due to
aberration and projection effects.  Similarly, the ordered magnetic
field in the material surrounding the jet results in an
strongly-ordered $\RM$-map, exhibiting nearly linear transverse
gradients along the length of the jet, and peaking near the core
(lower-left panel).

These features, however, are somewhat obscured upon convolving with
an observing beam.  While generally the optically thick core is
unresolved, it dominates the spectral index within a beamwidth of the
core.  The large $\RM$s near the core result in substantial beam
depolarization, shifting the peak polarized flux away from the peak
total flux.  While the derotated polarization vectors retain their
order, the polarization map can be dramatically effected, exhibiting
the nearly linear gradients only when located more than a beamwidth
from the core.  This is primarily due to the first term in Equation
(\ref{eq:RMavg}), with the second term typically making $\sim10\%$
corrections.  While this suggests that the spectral index terms may be
ignored, it should be noted that they produce gradients in random
directions, and therefore directly obscure the transverse gradients
indicative of the ordered toroidal magnetic fields.

\subsection{Dependence of the $\RM$ upon the AGN Parameters}\label{sec:JRMM-AP}
Here we address the importance of the jet and black hole parameters
upon the $\RM$ maps.  These arise via a number of the facets of our
jet model, depending upon both the fundamental properties of the AGN
itself as well as the assumptions that enter into the extrapolation
scheme.

\subsubsection{Jet Orientation}
Chief among the jet properties affecting both the $\RM$ maps and the
emission is the orientation.  Having a self-consistent 3D-jet model
provides the unique opportunity to assess the systematic uncertainty
associated with non-axisymmetric jet structures, providing a
fundamental lower limit upon the uncertainty associated with efforts
to infer information about the circum-jet material from $\RM$
distributions.

\begin{figure*}
\begin{center}
\includegraphics[width=\textwidth]{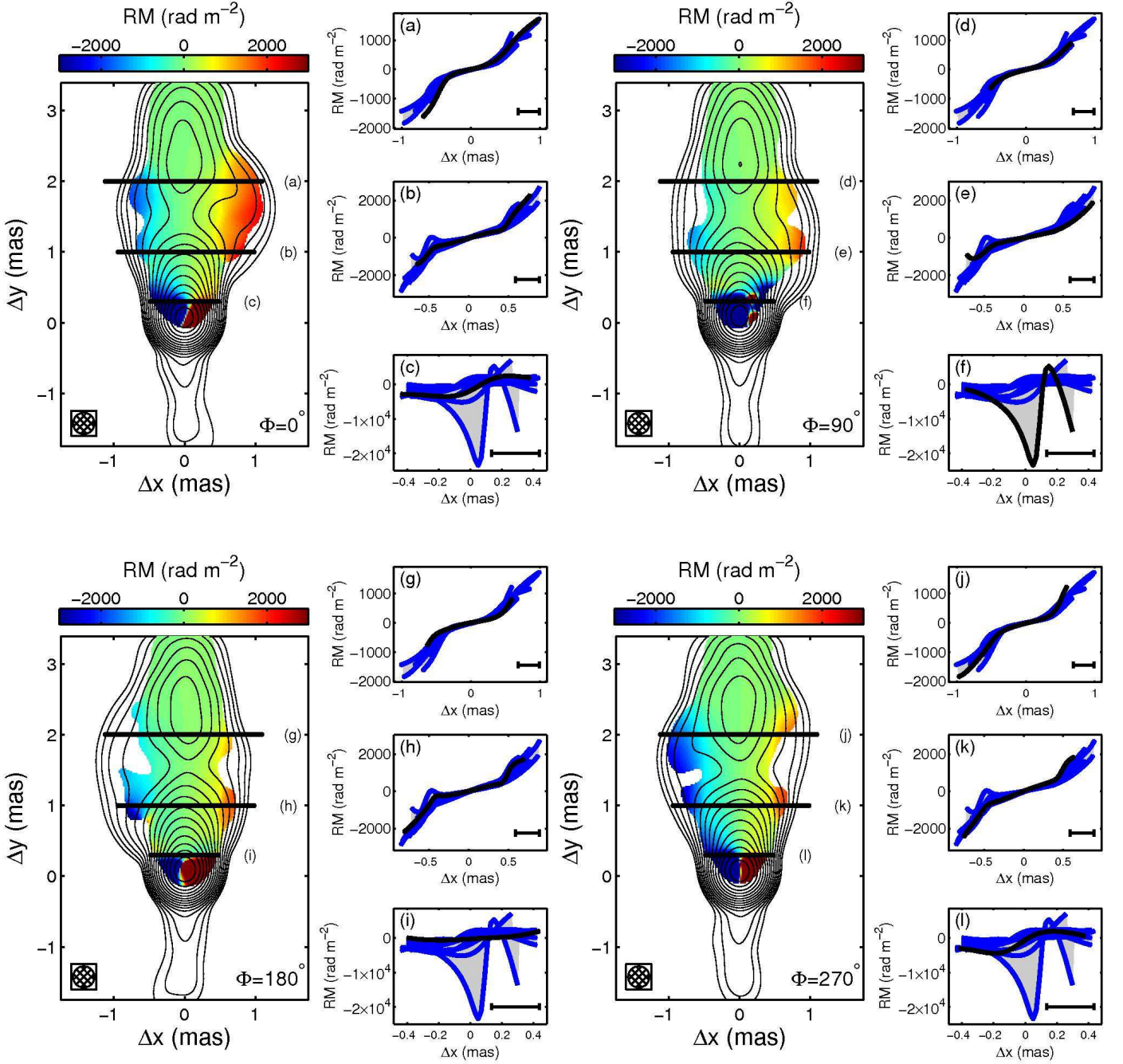}
\end{center}
\caption{$\RM$ maps at $15\,\GHz$ for our canonical model (see Table
  \ref{tab:simparams} for a complete list of relevant parameters), as
  seen from a number of azimuthal angles, $\Phi$ (listed in the
  lower-right of each $\RM$ map).  For reference total
  flux contours are overlayed, in logarithmic factors of 2, with the
  minimum contour corresponding to $0.98\,\mJy\,\mas^{-2}$.  $\RM$s are
  shown only where the polarized flux exceeds $1\mJy\,\mas^{-2}$.
  Transverse profiles, corresponding to the labeled black bars, are
  shown to the right of each $\RM$ map (note the differences in the
  angular ranges).  In addition to the $\RM$ profile for the view
  shown in each map (black), those from views with $\Phi=0^\circ$,
  $45^\circ$, $90^\circ$, $135^\circ$, $180^\circ$, $225^\circ$,
  $270^\circ$ and $315^\circ$ are shown in blue for comparison (with
  the enclosed envelope shown in grey).  In all maps and transverse
  sections the beam size is shown in the lower-left and lower-right,
  respectively.
  \label{fig:azi}}
\end{figure*}
Figure \ref{fig:azi} shows the $\RM$ maps for our canonical jet model
as viewed from a variety of azimuthal angles, $\Phi$ (though all at
the identical inclinations).  To the right of each $\RM$ map we show
example transverse profiles, taken near the core, far from the core
and at an intermediate location.  Within these the transverse
profiles for views with $\Phi=0^\circ$, $45^\circ$, $90^\circ$,
$135^\circ$, $180^\circ$, $225^\circ$, $270^\circ$ and $315^\circ$ are
shown in addition to that associated with the neighboring map,
providing a sense of the variations along each transverse section.
Large variations among views indicates large inherent uncertainties
in the $\RM$ values and their gradients due to the 3D jet structure.

Qualitatively, all azimuthal views are similar, exhibiting substantial
beam depolarization in the radio core, and a region of large $\RM$s
near the polarized flux maximum.  Despite being nearly linear at all
locations shown prior to beam convolution, where the transverse jet
structure is not resolved the $\RM$ profiles vary significantly.
However, despite the large structural variations in the jet itself, at
positions further than a single beam, the transverse gradients become
much more robust and are indicative of those present at ideal
resolution.

Note that while the presence of substantial variations between
transverse $\RM$ profiles does directly imply large uncertainties in
the $\RM$ gradients, the inverse is not true.  That is, consistent
$\RM$ gradients do not necessarily imply that these are indicative
of the circum-jet structure.  As implied by Equation (\ref{eq:RMgrad})
and as we shall see explicitly below, it is
possible for large beams to artificially produce essentially linear
gradients that are considerably smaller than the associated intrinsic
values.

\begin{figure}
\begin{center}
\includegraphics[width=\columnwidth]{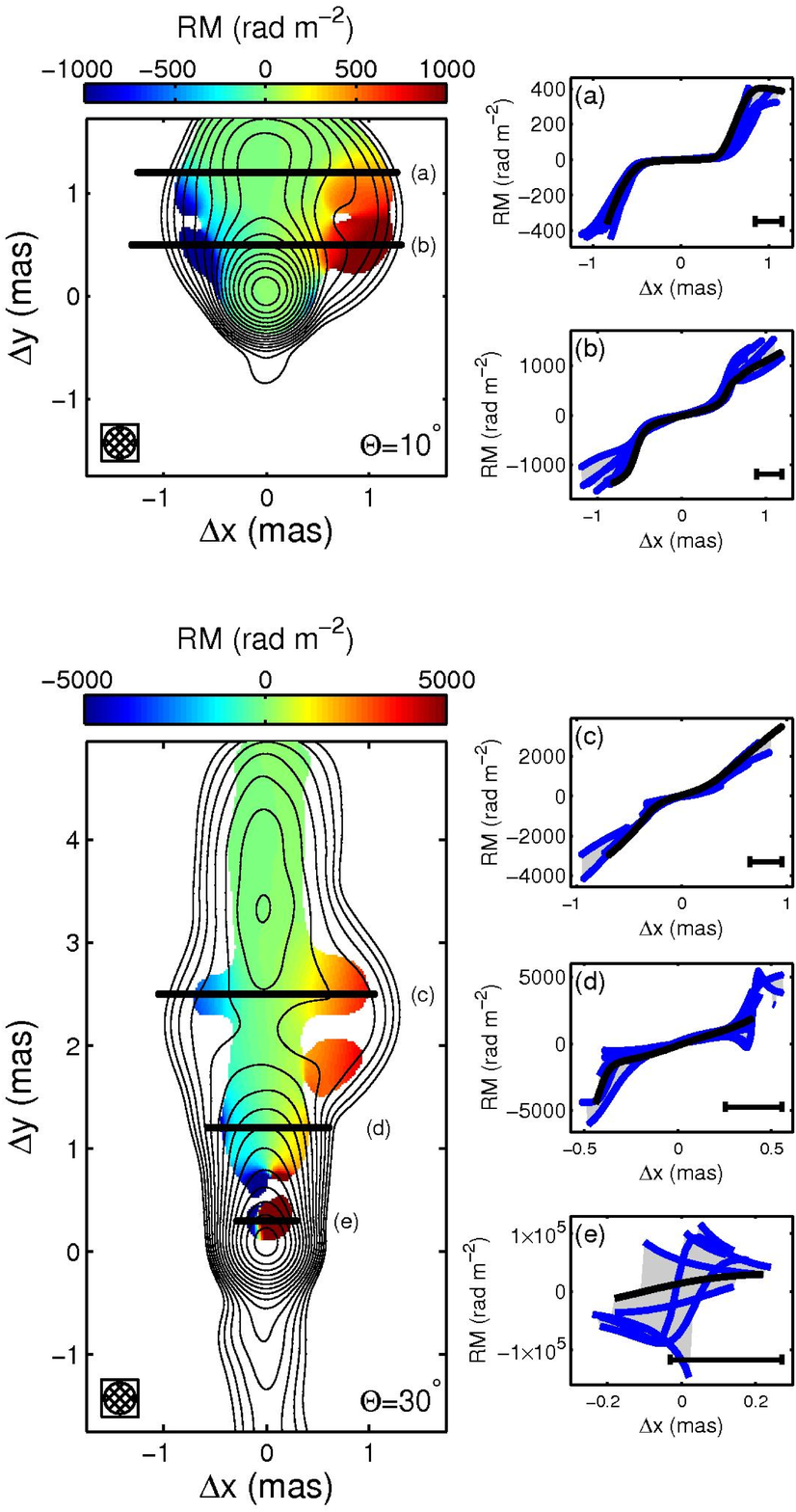}
\end{center}
\caption{$\RM$ maps at $15\,\GHz$ for jets viewed at inclinations of
  $10^\circ$ (top) and
  $30^\circ$ (bottom; \cf the upper-left panels of Figure \ref{fig:azi}; see
  Table \ref{tab:simparams} for a complete list of relevant parameters).
  For reference total flux
  contours are overlayed, in logarithmic factors of 2, with the
  minimum contour corresponding to $0.98\,\mJy\,\mas^{-2}$.  $\RM$s are
  shown only where the polarized flux exceeds $1\mJy\,\mas^{-2}$.
  To the right
  of each $\RM$ map,
  illustrative transverse $\RM$-profiles, corresponding to the
  labeled bars are shown (black) along with those from a variety of
  azimuthal viewing angles (blue).  In all maps and transverse
  sections the beam size is shown in the lower-left and lower-right,
  respectively.
  \label{fig:inc}}
\end{figure}
Jet inclination is known to have dramatic consequences for the
appearance of AGN jets.  This is also true for jet $\RM$-maps.
Generally, where the emission becomes
increasingly core dominated the finite beam effects (see Section
\ref{sec:JRMM-OP}) are exacerbated.  In addition, at different jet
inclinations the $\RM$s probe different circum-jet regions, with
smaller $\Theta$ being sensitive to larger radii.

In Figure \ref{fig:inc} we show the jets with inclinations of
$\Theta=10^\circ$ and $30^\circ$.  In order to ensure we are comparing
observationally similar systems, we adjust the accretion rate so that
the total flux is comparable to our canonical model, requiring
variations of a factor of two in each direction (see Table
\ref{tab:simparams}).  With this constraint, the $\RM$s are typically
smaller at smaller inclinations, typically resulting in less
depolarization.  Indeed, at $\Theta=10^\circ$ the radio core is not
depolarized, though finite beam effects do washout the $\RM$ gradients
at this location.  As with our canonical model, beyond a single beam
width the transverse $\RM$ profiles become indicative of the ordered,
ideal resolution values.

At larger inclinations the typical $\RM$s increase.  As in the
canonical case, the transverse $\RM$ profiles across the polarized flux
core exhibit large variations among azimuthal viewing angles, and
are thus unreliable measures of the circum-jet material.  Less obvious
is the fact that the intermediate transverse profile, Figure
\ref{fig:inc}d, also fails to reproduce the ideal resolution $\RM$
gradients, despite the strong correlation between $\RM$ profiles
obtained from differing azimuthal viewing directions.  Based upon the
simple model for beam-broadened $\RM$ gradients in Equation
(\ref{eq:RMgrad}), we would anticipate a gradient roughly four times
larger.  Thus the substantially reduced values of the $\RM$ gradient
cannot be accounted for simply due to the beam-broadened underlying
$\RM$ gradients.  Nevertheless,
at sufficiently large distances, where the jet is resolved in the
transverse direction, the $\RM$ gradients are reproduced with high
fidelity, and the large-scale structure is clearly evident.

\subsubsection{Black Hole Mass}
\begin{figure*}
\begin{center}
\includegraphics[width=\textwidth]{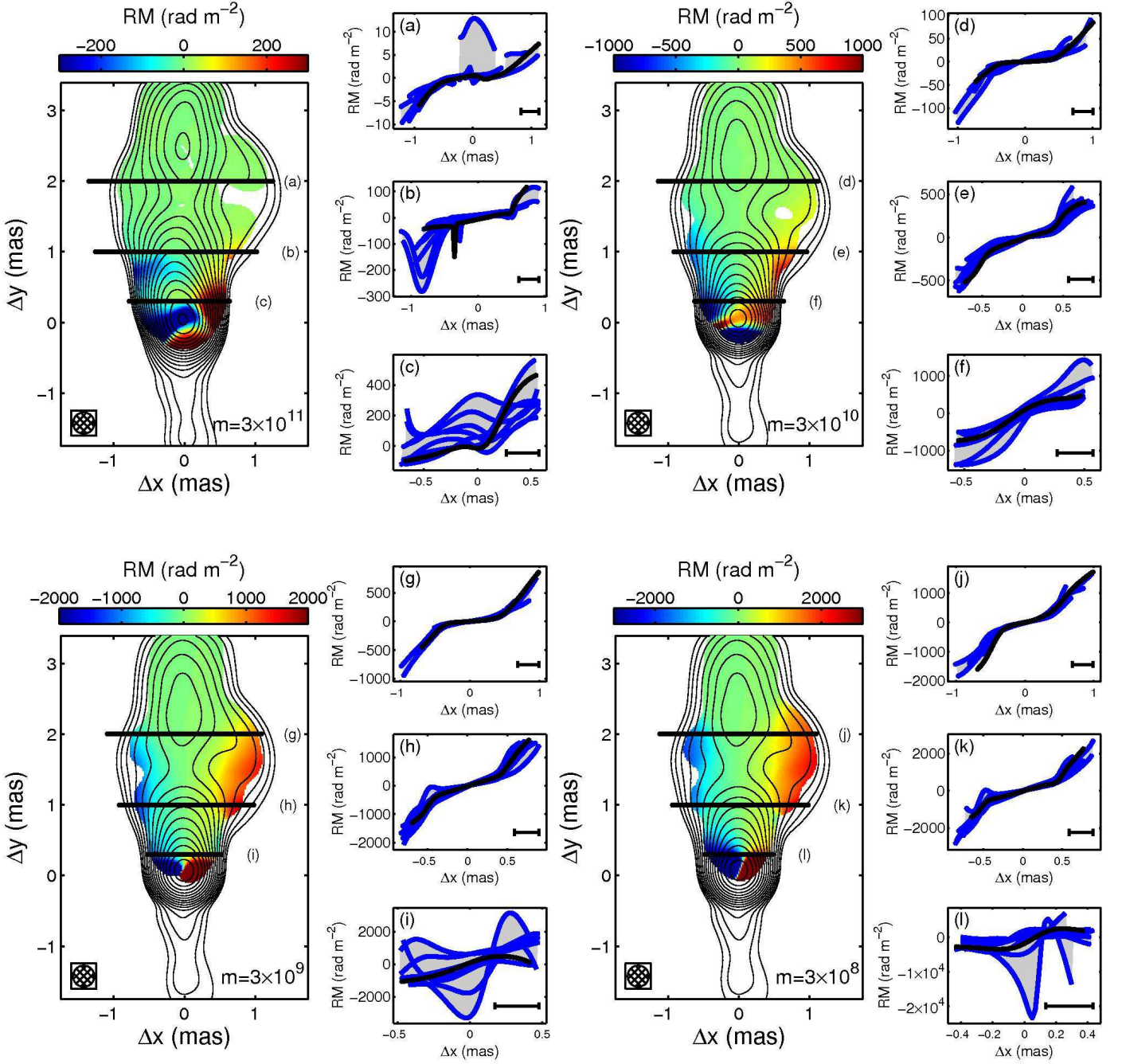}
\end{center}
\caption{$\RM$ maps at $15\,\GHz$ for jets produced by black holes
  with a variety of masses ($m\equiv M/\Ms$ is listed in the lower-right of
  each $\RM$ map; see Table \ref{tab:simparams} for a
  complete list of relevant parameters). 
  Of particular note is the fact that at $3\times10^{11}\Ms$ no
  extrapolation is required.  Our canonical model is shown in the
  lower-left.
  For reference total flux
  contours are overlayed, in logarithmic factors of 2, with the
  minimum contour corresponding to $0.98\,\mJy\,\mas^{-2}$.  $\RM$s are
  shown only where the polarized flux exceeds $1\mJy\,\mas^{-2}$.
  To the right of each $\RM$ map,
  illustrative transverse $\RM$-profiles, corresponding to the
  labeled bars are shown (black) along with those from a variety of
  azimuthal viewing angles (blue).  In all maps and transverse
  sections the beam size is shown in the lower-left and lower-right,
  respectively.
 \label{fig:mass}}
\end{figure*}
Variations in the black hole mass serve both to assess the importance
of the intrinsic astrophysical variation in the masses of AGNs as well
as the effects of our extrapolation scheme (Section \ref{sec:CM:ex}).
Here we choose to keep the physical size of the region of interest
fixed, and therefore change $\Rmax/\Rsim\propto M^{-1}$, i.e., change
the degree of extrapolation.  Since in all cases we are looking at the
same underlying jet structure, we are necessarily gauging the relative
importance of the black hole mass, or more specifically the length
scale it provides.

Again we must address the problem of constructing observationally
comparable systems.  This is complicated by the fact that
$m\equiv M/\Ms=3\times10^{11}$ systems, the mass required to avoid any
extrapolation given the physical size ($15\,\pc$) and distance
($300\,\Mpc$) under consideration, are physically unrealistic.  In
principle, as in Figure \ref{fig:inc} we would like to set
$\dot{M}$ such that the total fluxes are comparable.  However, this
suggests the natural choice of fixed $\dot{M}$, or
$\dot{M}/\dot{M}_{\rm Edd}\propto M^{-1}$, which is what we choose.
If the emission were optically thin everywhere this would give the
desired result.  In practice the total flux varies by a factor of 3
over variations in mass over 3 orders of magnitude, ostensibly due to
the subtlety different magnetic field geometry (see below).  We might
also expect the $\RM$s to be comparable: with $n\propto\dot{M}/r^2$
and $B\propto \left.n\right|_{r=M}^{1/2}(M/r)\propto\dot{M}^{1/2}/r$ we
have $\RM \sim n B r \propto \dot{M}^{3/2}/r^3$ and thus at fixed $r$,
$\RM\sim\dot{M}^{3/2}$.  This is not, however, the case, with the
typical $\RM$s of less massive black holes being considerably larger
than those of their larger analogs.  Far from the radio core, where
the beam convolved $\RM$s are indicative of the true values, we find
that in practice $\RM\propto M^{-1}$  (though, for a given mass the
$\RM$s do have the expected dependence upon accretion rate).

Near the black hole the toroidal and poloidal components of the
magnetic field have similar magnitudes.  Thus the results of
Section \ref{sec:CM:ex} imply that at fixed physical distance
$b^P/b^\phi \propto M$.  Since outside the jet $\bmath{b^P}$ undergoes
stochastic variations, we may therefore expect to find less ordered $\RM$
profiles at larger masses, exhibiting more vertical structure.  As
seen in Figure \ref{fig:mass}, this is indeed the case.  For
$m\equiv M/\Ms \gtrsim 3\times10^{10}$ vertical gradients can begin to
dominate the $\RM$ map near the core, extending further and becoming
more prominent as $m$ increases.  This vertical structure is an
intrinsic effect, appearing in the ideal resolution $\RM$ maps as
well.  This is in contrast to $\RM$-gradient reversals near the core
for $m\lesssim3\times10^{10}$, which are exclusively due to finite beam
effects.

The transverse gradients near the radio core at lower masses are
clearly not indicative of the true, nearly linear $\RM$ gradients.
However, for $m\gtrsim3\times10^{10}$ the transverse $\RM$ profiles in
Figure \ref{fig:mass}c \& \ref{fig:mass}f are more robust, exhibiting
a clearly identifiable gradients.  While the $\RM$ gradients along
the nearby sections are well below the true values, they are
consistent with the beam-broadened values anticipated by Equation
(\ref{eq:RMgrad}).  Since in practice the jet width is not known
a priori, determining the true gradients along these sections is
not possible for AGN jets associated with the most massive black holes
expected in nature.  Nevertheless, in this case the presence of a
robust gradient, even if unresolved, is indicative of ordered toroidal
fields.

\subsubsection{Magnetic Dissipation in the Faraday Screen}
\begin{figure*}
\begin{center}
\includegraphics[width=\textwidth]{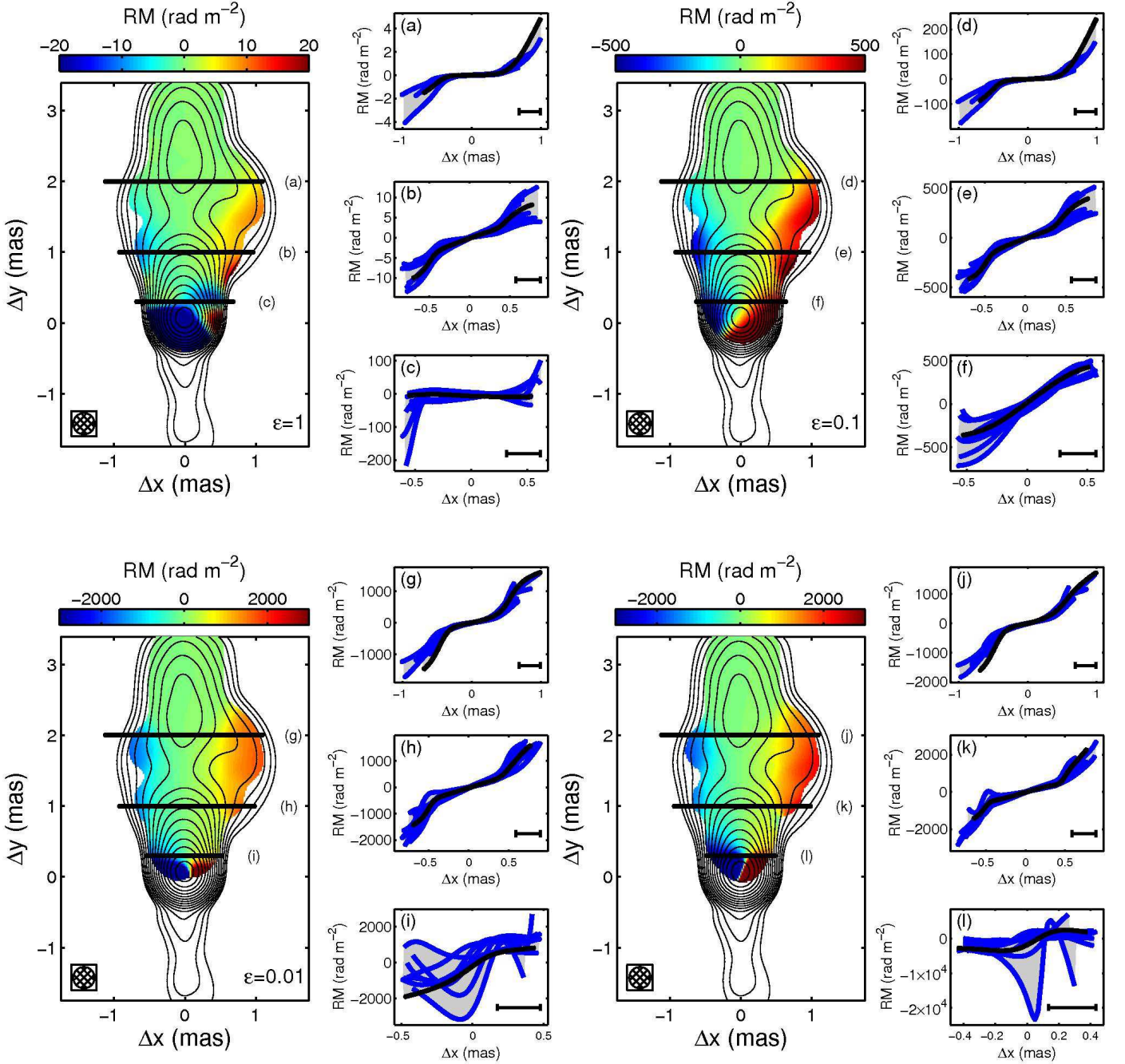}
\end{center}
\caption{$\RM$ maps at $15\,\GHz$ for jets in which the jet sheath is
  efficiently heated via magnetic dissipation ($\varepsilon$ is listed in
  the lower-right of each $\RM$ map; see Table
  \ref{tab:simparams} for a complete list of relevant parameters).
  For each the internal
  energy is extrapolated
  using $a_u=-1.3$ until it reaches $\varepsilon b^2/8\pi$ (generally,
  this occurs quite close to the black hole).  The upper-left,
  upper-right and lower-left panels correspond to $\varepsilon=1.0$, $0.1$
  and $0.01$, respectively.  For comparison the canonical model
  ($a_u=-8/3$) is shown in the lower-right panel.
  For reference total flux
  contours are overlayed, in logarithmic factors of 2, with the
  minimum contour corresponding to $0.98\,\mJy\,\mas^{-2}$.  $\RM$s are
  shown only where the polarized flux exceeds $1\mJy\,\mas^{-2}$.
  To the right of each $\RM$ map,
  illustrative transverse $\RM$-profiles, corresponding to the
  labeled bars are shown (black) along with those from a variety of
  azimuthal viewing angles (blue).  In all maps and transverse
  sections the beam size is shown in the lower-left and lower-right,
  respectively.
\label{fig:diss}}
\end{figure*}
Finally, we consider the dependence of the $\RM$s upon the rate at
which magnetic dissipation heats the circum-jet material.  The
transition from hot ($kT>m_e c^2$) gas within the jet to cool
($kT<m_e c^2$) gas surrounding the jet is typically rapid, occurring
over less than a jet width.  Nevertheless, since $f(T)$ is such a
strong function of $kT/m_e c^2$ (see Equation (\ref{eq:RMavg})), the
location of this transition can substantially affect the typical
$\RM$s produced near the jet.  Since $b^2/8\pi\rho c^2 > 1$ within
the jet, efficient dissipation of magnetic energy in the jet sheath
can easily heat the circum-jet material sufficient to suppress Faraday
rotation.

This is explicitly seen in Figure \ref{fig:diss}, which
shows the efficient heating model for a number of maximum heating
rates $\varepsilon$ (recall, we set the maximum internal energy to
$\varepsilon b^2/8\pi$).  When $\varepsilon=1.0$ the typical $\RM$s are
reduced by more than two orders of magnitude.  In contrast, setting
$\varepsilon\lesssim0.01$ results in nearly identical values to those
from our canonical model.  In all cases, with the exception of regions
near the core, the basic structure of the $\RM$ maps and their
transverse profiles are qualitatively similar.

As for the canonical case (shown in the lower-right panel of
Figure \ref{fig:diss}), where the transverse structure of the jet is
resolved the $\RM$s are indicative of the true values.  Similarly,
the transverse profiles near the core are typically unreliable.  An
exception is the appearance of a robust gradient at this location when
$\varepsilon=0.1$ (Figure \ref{fig:diss}f).  In this case the resulting
beam-convolved gradients are indeed consistent with those expected
from Equation (\ref{eq:RMgrad}).  In contrast, despite exhibiting a
general trend, the $\RM$s in Figure \ref{fig:diss}i are roughly an
order of magnitude lower than would be expected.  As a result, it may
be difficult to assess the fidelity of an observed transverse $\RM$
gradient when the jet width is insufficiently unresolved.

\subsection{Dependence of the $\RM$ upon the Observation Parameters}\label{sec:JRMM-OP}
In addition to the parameters of the AGN model, the measured $\RM$ can
depend upon how they are observed as well.  Here we address two ways
in which this occurs: observing frequency and beam size.  In order to
isolate the effects of each, we treat them separately by fixing one while
varying the other.  All of these are essentially finite beam effects;
in all cases shown in Figures \ref{fig:freq} \& \ref{fig:beam} the
intrinsic, ideal resolution $\RM$ maps are identical.

\subsubsection{Beam Size}
\begin{figure*}
\begin{center}
\includegraphics[width=\textwidth]{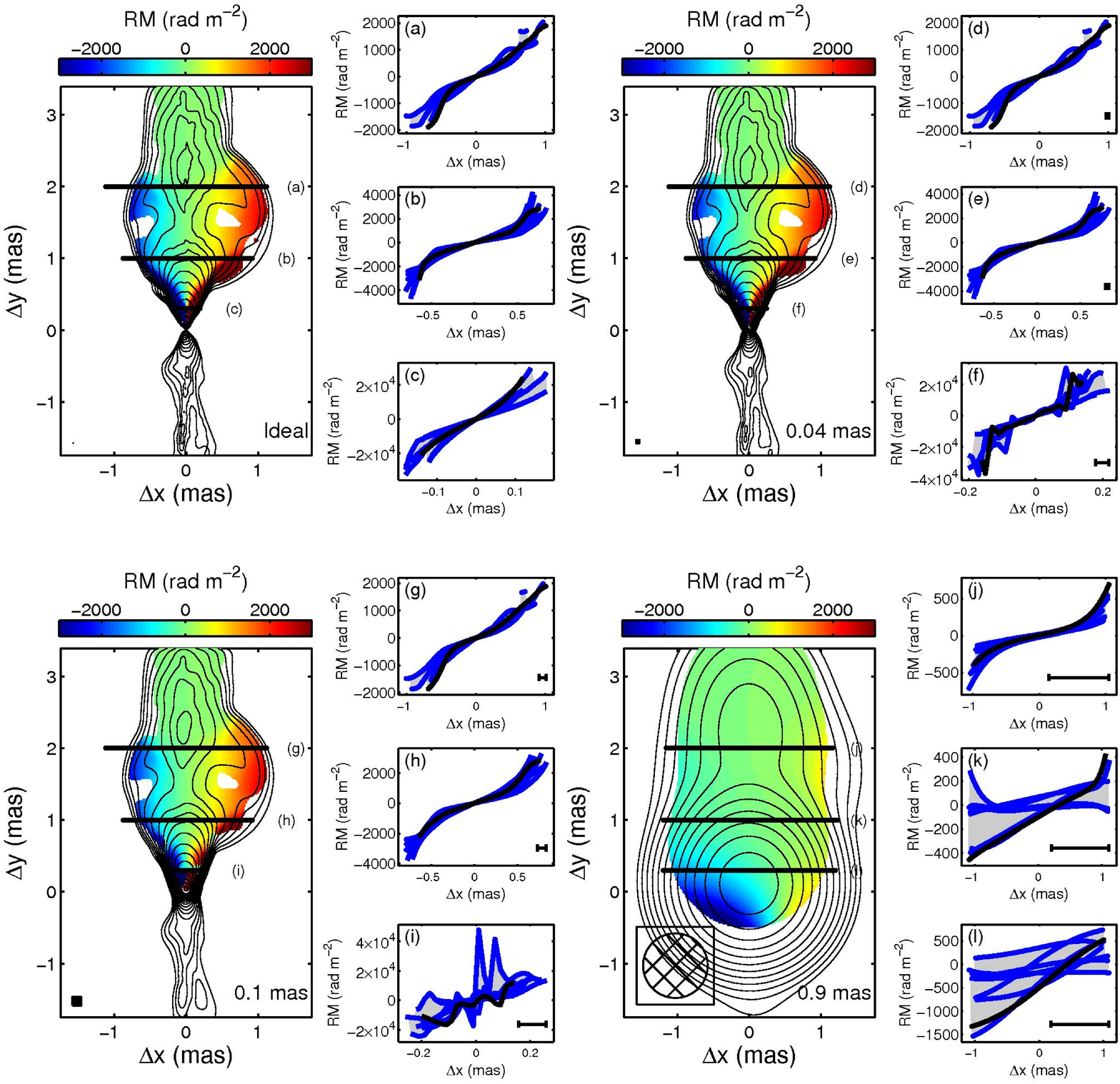}
\end{center}
\caption{$\RM$ maps of our canonical model at $15\,\GHz$ as observed
  with a number of 
  different beam sizes (beam FWHMs are listed in the lower-right of
  each $\RM$ map; cf. the upper-left panel of Figure \ref{fig:azi};
  see Table \ref{tab:simparams} for a complete list of relevant
  parameters).  In particular, we show the map at the computed pixel
  resolution ($0.02\,\mas$, upper-left), the highest design resolution
  of \VSOPII~($0.04\,\mas$, upper-right), the highest typical VLBA
  resolution ($0.1\,\mas$, lower-left) and the typical VLBA resolution
  at $5\,\GHz$ ($0.9\,\mas$, lower-right).
  For reference total flux
  contours are overlayed, in logarithmic factors of 2, with the
  minimum contour corresponding to $0.98\,\mJy\,\mas^{-2}$.  $\RM$s are
  shown only where the polarized flux exceeds $1\mJy\,\mas^{-2}$.
  To the right of each $\RM$ map,
  illustrative transverse $\RM$-profiles, corresponding to the
  labeled bars are shown (black) along with those from a variety of
  azimuthal viewing angles (blue).  In all maps and transverse
  sections the beam size is shown in the lower-left and lower-right,
  respectively.
 \label{fig:beam}}
\end{figure*}
We have alluded to the importance of beam size a number of times
already; generally finding that where the beam does not resolve the
transverse structure of the jet the transverse $\RM$ profiles are
inaccurate.  Here and in Figure \ref{fig:beam}  we make these
statements explicit by convolving the same image (corresponding to our
canonical model; see Table \ref{tab:simparams} for details) with a
number of different beams.  These correspond to the highest typical
VLBA resolution at $5\,\GHz$ ($0.9\,\mas$), highest typical VLBA
resolution at $43\,\GHz$ ($0.1\,\mas$), highest design resolution for
VLBI using the upcoming space-based radio telescope
\VSOPII~($43\,\GHz$, $0.04\,\mas$, \citealt{VSOP2:09}) and the computational resolution
($\sim0.02\,\mas$).
These can also be compared with our canonical model
(e.g. Figure \ref{fig:azi}), which has a beam typical of VLBA
observations at $15\,\GHz$.

The $0.9\,\mas$ and $0.3\,\mas$ beams are typical
of matched-resolution $\RM$ studies, and thus provide insight into the
ability of the existing literature to constrain the parameters of
simulated jets.  \VSOPII~promises to achieve beam sizes of $\sim0.1\,\mas$
at $8\,\GHz$, roughly a factor of two improvement over the VLBA.
Thus, $0.1\,\mas$ beam is indicative of what may be possible in the
near future.  Finally, \VSOPII's maximum resolution gives a
lower-limit on realistically attainable beam widths over the next
decade, though in this case the observed $\RM$s would be obtained
only over a relatively small wavelength range.

As previously stated, the ideal resolution case produces clear, nearly
linear transverse gradients along the length of the jet.  \VSOPII~
should be able to probe gradients as close as $0.3\,\mas$ to the radio
core.  Similarly, high frequency observations at the VLBA faithfully
reproduce the transverse $\RM$ gradients in the optically thin region,
though near the radio core large spurious fluctuations develop.

In contrast, for large beams, while apparently linear gradients exist,
these are in some cases clearly anomalous, arising from the breadth of
the beam instead of the structure of the jet.  This is most clearly
evident in the bottom two transverse profiles, Figure \ref{fig:beam}k
\& \ref{fig:beam}l, which show profiles with a number of different
gradients, in some cases with the wrong sign.  In Figure
\ref{fig:beam}k, the presence of inverted gradients provides a strong
case against interpreting these as due to the true structure in the
$\RM$ map.  However, the average profile in this case is roughly
consistent (within 50\%) with that anticipated by Equation
(\ref{eq:RMgrad}).  In contrast, Figure \ref{fig:beam}l exhibits $\RM$
gradients which are all of the correct sign, though also all
well below the values expected even after accounting for the
beam-broadening of the profiles.  Far from the radio core, despite
being unresolved, the transverse gradients are strongly correlated in
azimuthal viewing angle and are well-described Equation
(\ref{eq:RMgrad}), producing gradients roughly a factor of 4 smaller
than the true values.  Thus, despite the inability to infer the actual
$\RM$ values due to the failure to resolve the jet transverse
structure, the structure in the $\RM$ map is indeed indicative of the
ideal-resolution $\RM$s more than a beam width from the radio core.

\subsubsection{Frequency}
\begin{figure*}
\begin{center}
\includegraphics[width=\textwidth]{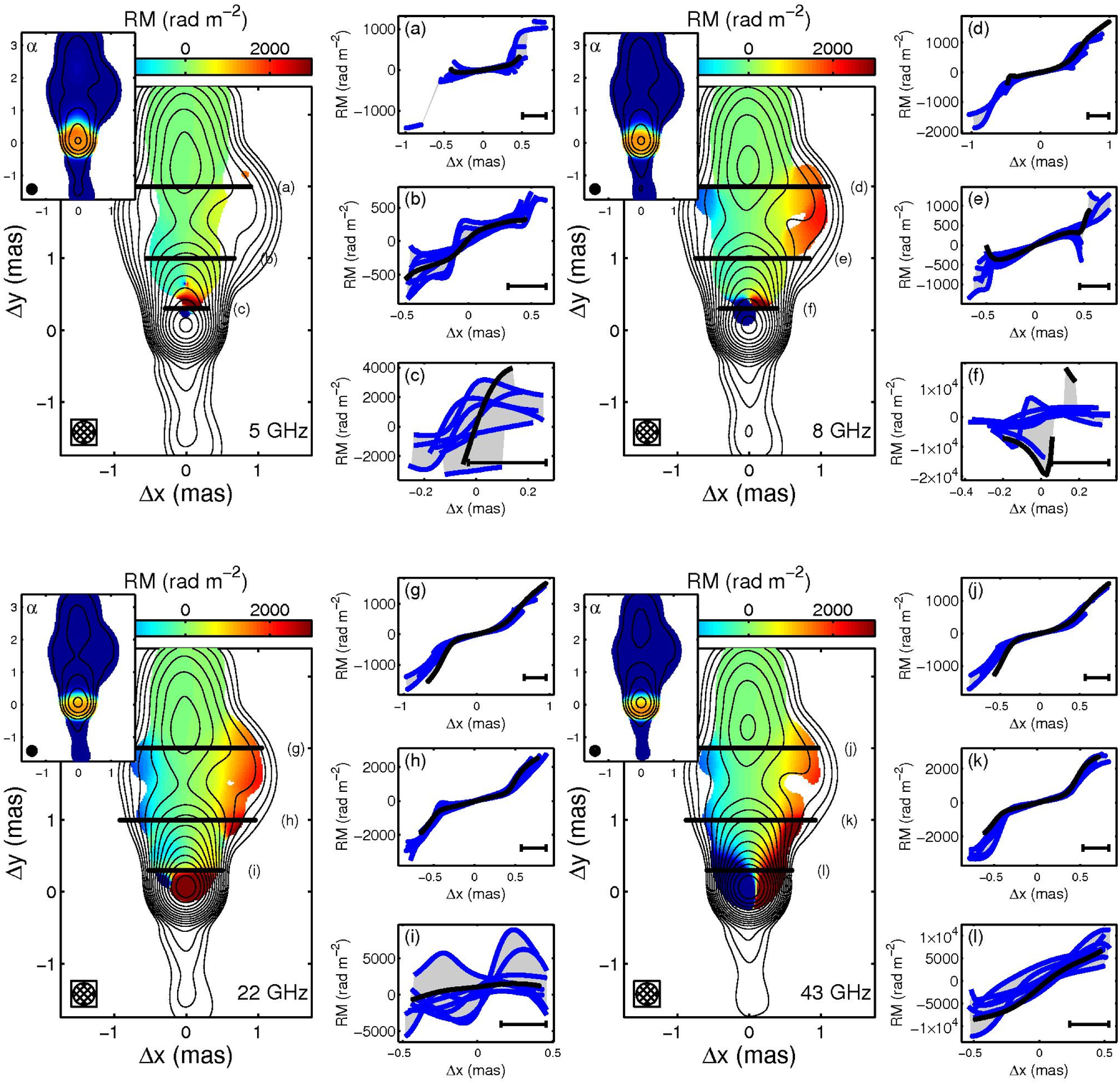}
\end{center}
\caption{$\RM$ maps of our canonical model as observed $\nu=5\,GHz$,
  $8\,\GHz$, $22\,\GHz$ and $43\,\GHz$ (observation frequecies are
  listed in the lower-right of each $\RM$ map; cf. the upper-left panel of
  Figure \ref{fig:azi}; see Table \ref{tab:simparams} for particular
  parameters).  In all cases the beam size is fixed to $0.3\,\mas$.
  The beam-convolved spectral index map is shown in the upper-left
  inset of each $\RM$ map, in which the colors range from $-0.5$ (dark
  blue) to $0.5$ (dark red), indicating the size of the photosphere at
  each frequency.   For reference total flux
  contours are overlayed, in logarithmic factors of 2 (factors of 8 in
  the $\alpha$ map insets), with the
  minimum contour corresponding to $0.98\,\mJy\,\mas^{-2}$.  $\RM$s are
  shown only where the polarized flux exceeds $1\mJy\,\mas^{-2}$.
  To the right of each $\RM$ map, illustrative
  transverse $\RM$-profiles, corresponding to the labeled bars are
  shown (black) along with those from a variety of azimuthal viewing
  angles (blue).  In all maps and transverse sections the beam size is
  shown in the lower-left and lower-right, respectively.
 \label{fig:freq}}
\end{figure*}
While we have defined the $\RM$ in the differential sense, where the
$\lambda^2$-law holds in the ideal-resolution $\RM$ maps, we might
anticipate that frequency effects may
be neglected.  However, as seen in Figure \ref{fig:freq}
(cf. \ref{fig:azi}), near the radio core this is not the case
generally.  Despite having identical intrinsic $\RM$s, differences in
the photosphere size and beam depolarization associated with different
observing frequencies can dramatically alter the beam-convolved $\RM$
maps.  Note that here we fix the beam size to $0.3\,\mas$ despite the
change in observing frequency.

Firstly, the increase in beam depolarization at longer wavelengths
severely restricts the region over which the $\RM$s may be measured.
At $5\,\GHz$, the jet is a few beams across, and begins to suffer from
inadequate transverse resolution.  As a consequence, even where the
jet is optically thin, the beam-convolved $\RM$ map exhibits
substantial variations about the true values.  The $\RM$ profile shown
in Figure \ref{fig:freq}b is roughly a factor of two below the true
values, though in agreement with those predicted by Equation
(\ref{eq:RMgrad}).  By $8\,\GHz$ the jet is sufficiently resolved
(given our polarized flux limit of $1\,\mJy\,\mas^{-2}$) that the
gradients in the optically thin portions of the jet (e.g. the top two
transverse profiles) reproduce the true gradients with high fidelity.

Secondly, the size of the radio photosphere shrinks as frequency
increases, with a corresponding decrease in the consequences of the
rapid variations in the spectral index.  At $5\,\GHz$ the photosphere
extends roughly $0.5\mas$, larger than the beam itself, while at
$43\,\GHz$ it is comparable to the resolution with which we computed
the intrinsic $\RM$ maps ($0.02\,\mas$).  The decrease in the
importance of the optically thick core is seen most dramatically in
the appearance of the consistent transverse gradients close to the
core, Figure \ref{fig:freq}l, which are roughly an order of magnitude
below the true values, consistent with Equation (\ref{eq:RMgrad}).

When the beam size is increased to $0.9\,\mas$ all of the profiles at
$5\,\GHz$ fail to reproduce even the beam-broadened $\RM$ gradients.
Nevertheless, above $8\,\GHz$, as in Figure \ref{fig:beam}j,
transverse profiles far from the core exhibit robust gradients
indicative of the expected values for transversely unresolved jets.
This is also true at intermediate distances from the radio core
(Figure \ref{fig:freq}h \& \ref{fig:freq}k) for frequencies above
$22\,\GHz$.  Finally, at $43\,\GHz$, the highest frequency we
consider, even the core $\RM$ profiles are well-defined and in
agreement with the values anticipated by Equation (\ref{eq:RMgrad}),
though roughly two order of magnitudes smaller than the true values.
Thus, for the purposes of detecting robust $\RM$ gradients associated
with poorly resolved structure within the Faraday screen, pushing to high
frequencies alone can partially ameliorate resolution issues (though,
as addressed explicitly in Section \ref{sec:loc}, it may be difficult
to distinguish between nearby and distant Faraday screens in this
case).

The differences between the $\RM$ maps near the radio core due to beam
convolution necessarily implies that the $\lambda^2$-law is violated
in this region.  On the other hand, when the transverse jet structure
is sufficiently resolved and the observed $\RM$s agree with the
intrinsic $\RM$s, as occurs far from the core in the optically thin region,
then similarities between the $\RM$s implies that the $\lambda^2$-law
holds.  These regions also differ in how well the beam-convolved $\RM$
maps reproduce the true values.  Hence, this suggests that at least
for our simulated AGN jets, the $\lambda^2$-law provides a rough
measure of the accuracy of the $\RM$ gradients when the $\RM$s are
sufficiently resolved\footnote{This conclusion in specific to our jet
  models and may not be generic.  For example, unresolved point
  sources are capable of producing $\RM$s which appear to follow the
  $\lambda^2$ though are erroneous.  This does not appear, however, to
  be the case for the highly-ordered structures that appear within our
  simulated jets.}.

\subsection{Dependence of the $\RM$ upon the Jet Structure}\label{sec:JRMM-JS}
The primary motivation for studying the $\RM$ distributions of AGN
jets has been to infer the structure and parameters of the Faraday
screen.  In the context of the 3D simulation, neither of these are
free parameters, and we have obtained a characteristic, bilateral
structure for the $\RM$ maps, presumably as a consequence of
large-scale structure within the Faraday screen.  Here we address how
these morphological features are related to the structural properties
of the Faraday screen itself.  We do this by artificially removing
components of the magnetic field \& velocity and recomputing the $\RM$
map, while using our canonical model to produce the jet emission.

\begin{figure}
\begin{center}
\includegraphics[width=\columnwidth]{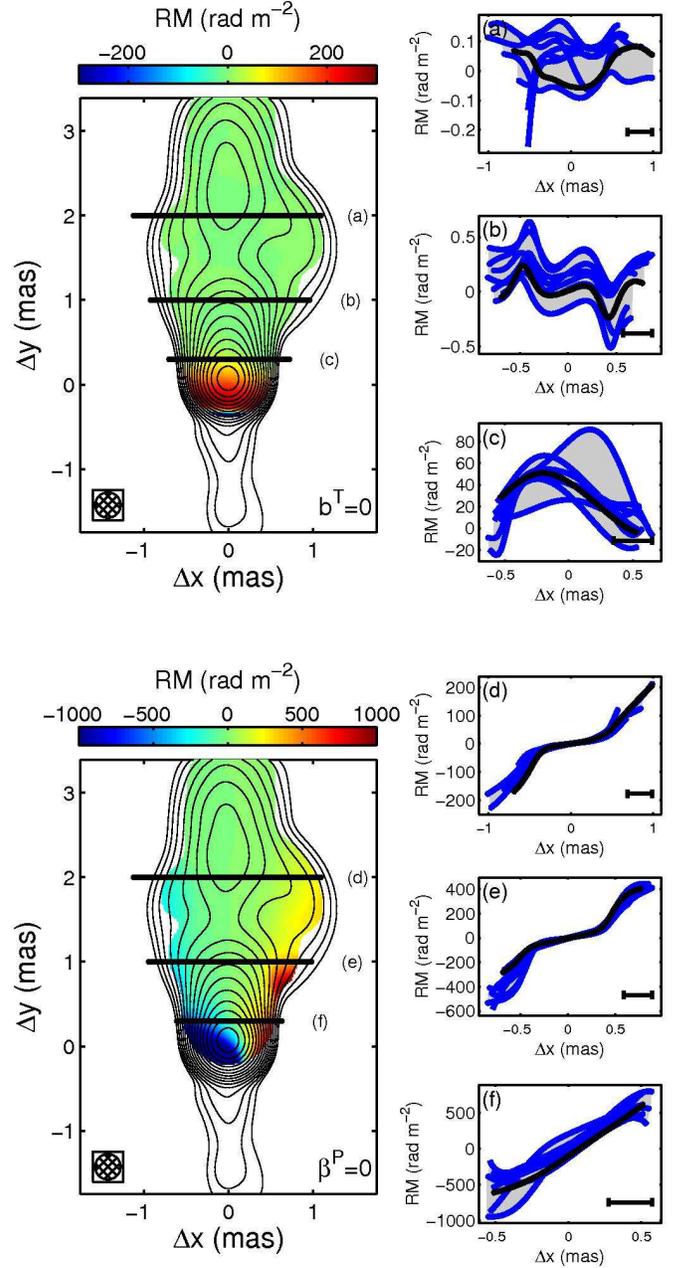}
\end{center}
\caption{$\RM$ maps at $15\,\GHz$ when toroidal component of the
  comoving magnetic field (top) and poloidal component of the velocity
  (bottom) of the plasma within the Faraday screen is made to
  artificially vanish (cf. the upper-left panel of Figure \ref{fig:azi};
  see Table \ref{tab:simparams} for a complete list of relevant
  parameters).
  For reference total flux
  contours are overlayed, in logarithmic factors of 2, with the
  minimum contour corresponding to $0.98\,\mJy\,\mas^{-2}$.  $\RM$s are
  shown only where the polarized flux exceeds $1\mJy\,\mas^{-2}$.
  To the right of each $\RM$ map,
  illustrative transverse $\RM$-profiles, corresponding to the
  labeled bars are shown (black) along with those from a variety of
  azimuthal viewing angles (blue).  In all maps and transverse
  sections the beam size is shown in the lower-left and lower-right,
  respectively.
\label{fig:struc}}
\end{figure}

\subsubsection{Magnetic Field Structure}
The relationship between the poloidal and toroidal magnetic field has
already been indirectly demonstrated to have a significant effect upon
the structure of the $\RM$ map.  Large poloidal fields (e.g., those
associated with $m=3\times10^{11}$) can result in the loss of
the transverse symmetry apparent outside of the radio core.  This is
illustrated explicitly in the top panel of Figure \ref{fig:struc}, which
presents the $\RM$ map when the toroidal component of the magnetic
field in our canonical model is artificially made to vanish.
Significant transverse gradients vanish completely everywhere within
the jet (where it is sufficiently resolved), confirming that
nearly-linear, resolved transverse gradients are due to large-scale,
ordered toroidal magnetic fields within the Faraday screen.

The remnant $\RM$ is that associated with the poloidal field, and is
thus roughly 3 orders of magnitude smaller than that in our canonical
model (see, Figure \ref{fig:azi}).  This is a function of the black
hole mass, with $b^P/b^T\propto M^{-1}$ at fixed physical distance,
and thus the perturbations due to poloidal fields will be larger for
more massive systems.  Nevertheless, it appears that for typical
blazars and BL Lac objects the fixed offset in the transverse profile
due to the presence of poloidal fields will be dominated by the
uncertainties in the location of the jet spine (not necessarily
located along the jet axis), viewing angle and finite beam effects for
single $\RM$ profiles.  It remains to be seen if it is possible to
reduce these uncertainties by combining multiple profiles, making use
of the strong correlation in these quantities along the jet axis.

\subsubsection{Velocity Structure}
The large Lorentz factors within AGN jets suggest that the velocity
structure of the Faraday screen may also be important.  The degree to
which this occurs in practice depends upon the location of the Faraday
screen and how coupled it is to the jet itself.

Setting the poloidal velocity to vanish, shown in the bottom panel of
Figure \ref{fig:struc}, made little difference to the overall
morphology of the $\RM$ map.  It does, however, decrease the typical
$\RM$s by a factor of roughly 4.  This is not unexpected if the
typical Lorentz factors within the Faraday screen are on the order of
2 (a fact that will be borne out in the following section); from
Equation (\ref{eq:RMavg}),
\begin{equation}
\frac{\left.\RM\right|_{\beta^P=0}}{\RM}
\simeq
\Gamma^2 \left(1-\beta^P\cos\Theta\right)^2
\simeq
\Gamma^2 \left[\frac{1}{\Gamma^2}+\frac{\Theta^2}{2}\right]^2\,,
\end{equation}
which when $\Theta\lesssim\sqrt{2}/\Gamma$ reduces to $\Gamma^{-2}$.
Unlike the canonical case, when the poloidal velocity is made to
vanish the transverse $\RM$ profiles exhibit a well-defined, nearly
linear gradients near the core.  These are roughly an order of
magnitude smaller than the true $\RM$s, though in agreement with
those expected by Equation (\ref{eq:RMgrad}).

In contrast, the toroidal velocities appear to have little effect
beyond the radio core.  This is not particularly surprising since from
Section \ref{sec:CM:ex} we have that
$\beta^T/\beta^P\sim\left(GM/c^2r\right)^{-1}$, and thus at the large
distances (in terms of the gravitational radius of the black hole)
seen here we anticipate the toroidal velocities to be roughly 5 orders
of magnitude smaller than the poloidal values within the screen.  Thus
the kinds of effects associated with large-scale relativistic motions
described in \citet{Brod-Loeb:09b} are likely to be unimportant in the
case of isolated, roughly stable jets.

\subsection{Location and Structure of the Jet Faraday Screen}\label{sec:loc}
While we have described how some of the most commonly discussed
features of the circum-jet environment imprint themselves upon the
$\RM$ map, we have said little about the location and structure of the
Faraday rotating medium.  Here we address this, identifying what
portion of the outflow is responsible for generating the Faraday
rotation and how it is related to the jet itself.  In addition, we
address how this may be distinguished from non-local models of the Faraday
screen.

\begin{figure}
\begin{center}
\includegraphics[width=\columnwidth]{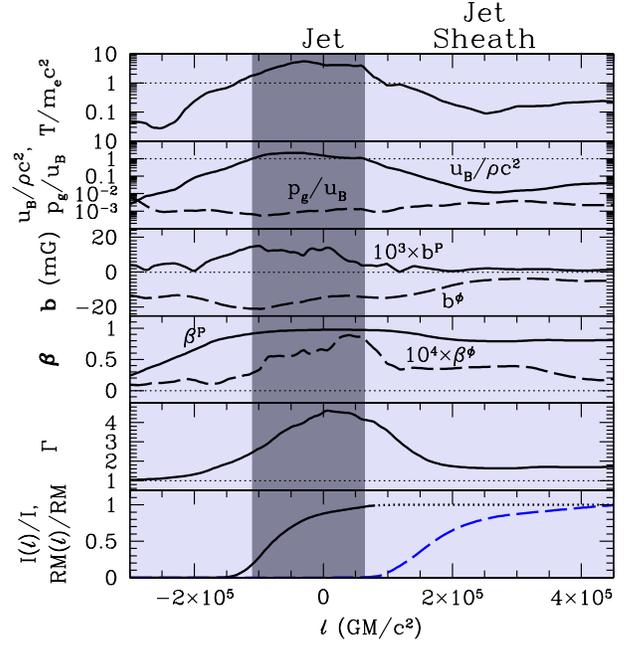}
\end{center}
\caption{Plasma quantities along an example line of sight, in
  comparison to the location of the Faraday screen.  The locations of
  the jet, defined by $\rho c^2 < u_B\equiv b^2/8\pi$, and its
  surrounding sheath, defined by $p_g<u_B/3$, are represented by the
  dark and light gray shaded regions, respectively.  This corresponds
  a line of sight responsible for producing the $\RM$ at an angular
  position of $\Delta\Omega_x=1\,\mas$ and
  $\Delta\Omega_y=1.5\,\mas$, and is the 
  second from the bottom line of sight shown in Figure
  \ref{fig:IRM}, with the intensity (solid black line) and rotation
  measure (dashed, blue line) shown explicitly in the bottom panel.
  \label{fig:raysec}}
\end{figure}
A number of relevant plasma quantities, in addition to $I_\nu$ and the
$\RM$, along a representative line of sight are shown in Figure
\ref{fig:raysec}, including the poloidal and toroidal components of
$\bmath{\beta}$ and $\bmath{b}$, the Lorentz factor and temperature.
Generally, the Faraday screen is quite close to the jet, lying within
two jet widths of the jet axis.  The bulk of the Faraday rotation
occurs within the strongly magnetized jet sheath, defined here by
when the gas pressure is less than $1/3$ of the magnetic pressure
(i.e., the plasma $\beta<1/3$), though at larger inclinations a
significant fraction may occur within the surrounding wind.  While the
Faraday screen and emission region are well-separated, they are both
parts of a larger MHD structure.  As illustrated in Figure
\ref{fig:raysec}, the jet sheath is a smooth extension of the jet
itself; i.e., there is no well-defined spine-sheath structure despite
our identification of a jet ``sheath'' \citep[cf.][]{lb02a,lb02b}.

The entire rotating region is unbound in the particle, thermal
{\em and} magneto-thermal senses, i.e.,
\begin{equation}
\begin{gathered}
-u_t \simeq \left(1-\frac{2GM}{c^2r}\right)\Gamma > 1\,,\\
-h u_t \simeq \left(1-\frac{2GM}{c^2r}\right) \frac{\rho+u_g+p_g}{\rho}\Gamma > 1\,,\\
-\frac{T^P_t}{\rho u^P} \simeq
\left(1-\frac{2GM}{c^2r}\right)
\frac{
\left(\rho+u_g+p_g\right)\Gamma^2\beta^P
-
b^P\bmath{\beta}\cdot\bmath{b}
}{\rho\Gamma\beta^P} > 1\,,
\end{gathered}
\end{equation}
respectively,
where $T^P_t$ is the contravariant poloidal, covariant time component
of the ideal MHD stress-energy tensor,
$u_t$ is the covariant time component of the 4-velocity,
and $u^P$ is the contravariant poloidal component of the 4-velocity.
Therefore, the Faraday screen is clearly associated
with the large-scale outflow.  Despite this, the Lorentz
factor of the Faraday rotating material is generally much lower than
the jet interior,
$\Gamma\simeq2$, due in part to the strong dependence upon $\Gamma$
of the integrand in Equation (\ref{eq:RMdef}).

The $\RM$ is dominated by contributions close to the jet
and not by the disk wind.
In the simulations, the disk wind is not as well-collimated as the jet
due to the finite extent of the accretion flow.
While the highly relativistic jet becomes mostly ballistic
when losing lateral support due to the finite disk extent,
the weakly relativistic disk wind fills the space as a more spherical outflow.
Even the part of the disk wind launched from near the black hole
has a significant loss of toroidal field strength with radius
compared to the relativistic jet due to the overall expansion
of the disk wind.
The simulations suggest that a disk extent of order $\pc$-scales
may be required to have a well-collimated disk wind
that significantly contributed to the $\RM$ on $\pc$-scales.

Throughout the duration of the 3D GRMHD simulation we have used,
there is no evidence for the presence of reversed field polarity
as in the ``magnetic tower'' picture discussed by \citet{mgb09}.
They suggested that the time-dependent sign of the
$\RM$ in BL Lac object B1803+784 can be explained best by magnetic
tower models containing nested, reversed helical fields, in which the
inner field changes (in time) its degree of winding relative to the outer return field.
In the GRMHD simulations, at no point along the line
of sight does the toroidal component of the comoving or lab-frame
magnetic field reverse direction
It appears unlikely
that a significant fraction of the observed $\RM$ can be due to a
distant, reversed region, as in a disk wind.
Furthermore, such reversed magnetic tower geometries appear to be at
odds with jets launched by accretion disks, for which the
interior of the jet is universally more tightly wound due to
the monotonically decreasing radial Keplerian rotation profile.
One might be able to invoke a slowly rotating black hole
connected to time-varying launching points in the Keplerian disk.
However, this requires a slowly rotating black hole with $a/M\lesssim0.4$,
at which the total power of the disk wind may dominate that of the black hole jet \citep{mck05}
leaving the emission dominated by a quasi-spherical region instead of a collimated jet.
Sign changes could potentially be due to changes in
the overall polarity of a dipolar field near the black hole,
although this may be at odds with the long-term constancy of the sign of the $\RM$ gradient.
On the other hand, the presence of instantaneous erroneous $\RM$
gradients (even sign changes) due to the finite beam size
can produce qualitatively similar behaviors near the radio core,
and might be responsible for the apparent time-dependence, though we
should note that in B1803+784 the reversal occurred far from the radio
core.  An investigation of $\RM$s that considers these issues,
such as considering $\RM$s from time-dependent simulations,
is left for future work.

\begin{figure*}
\begin{center}
\includegraphics[width=\textwidth]{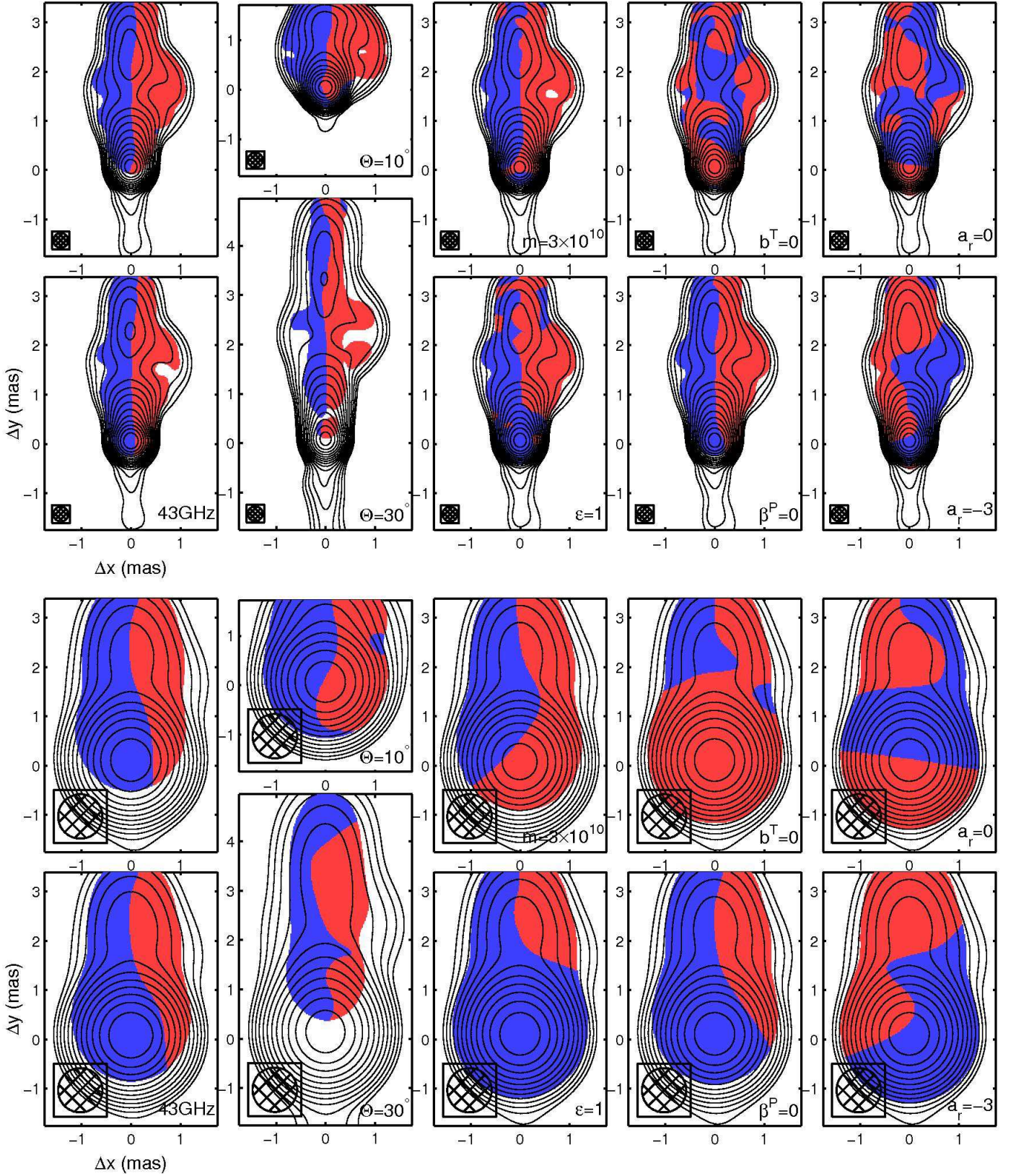}
\end{center}
\caption{$\sgn(\RM)$ for an illustrative subset of the cases discussed
  in the text, compared to two examples of random foreground screens
  (rightmost maps).  These are shown for two beam sizes, corresponding
  to resolved (top) and unresolved (bottom) jets.  See the text and Table
  \ref{tab:simparams} for a complete list of  relevant parameters for
  each, though the relevant defining parameter is listed as it appears
  in Figures \ref{fig:inc}--\ref{fig:struc} in the
  lower-right corner of each $\sgn(\RM)$ map.
  For reference total flux
  contours are overlayed, in logarithmic factors of 2, with the
  minimum contour corresponding to $0.98\,\mJy\,\mas^{-2}$.  $\RM$s are
  shown only where the polarized flux exceeds $1\mJy\,\mas^{-2}$.
  In all maps the beam size is shown in the lower-left corner.
  \label{fig:sRM2}}
\end{figure*}

The proximity of the Faraday screen to the jet makes a strong
prediction regarding the morphology of the $\RM$ maps.  The presence
of a large-scale radially correlated, toroidally dominated magnetic
field within the jet sheath implies that the $\RM$ map will be
strongly correlated along the jet.
This has
dramatic consequences for the $\sgn(\RM)$ maps, resulting in a very
particular bilateral structure.

Example $\sgn(\RM)$ maps are presented for a variety of jet model
and observational parameters in Figure \ref{fig:sRM2}.  For two
different beam widths, chosen to represent resolved and unresolved
jets, ($0.3\,\mas$ and $0.9\,\mas$ on the top and bottom,
respectively) we present the $\sgn(\RM)$ maps representative
of the various parameters we have considered in Sections
\ref{sec:JRMM-AP}--\ref{sec:JRMM-JS}.  Specifically, for each beam
size we show the $\sgn(\RM)$ associated with our canonical model
(upper left), at $43\,\GHz$ (lower left), high and low inclinations
(upper and lower center-left, respectively), a black hole with
$m=3\times10^{10}$ (upper center), an efficiently heated Faraday
screen (lower center), vanishing toroidal magnetic field (upper
center-right) and vanishing poloidal velocity (lower center-right).
The $\sgn(\RM)$ morphology is insensitive to the properties of the
poloidal magnetic field and the toroidal velocity, though both can
alter the details of the $\sgn(\RM)$ map.

With the exception of a vanishing toroidal magnetic field, all of the
jet models listed above exhibit the afore mentioned bilateral
morphology throughout the optically thin portion of the jet (as with
the $\RM$ profiles, near the radio core departures from this structure
can occur for some azimuthal orientations).  This suggests that if the
Faraday screen is close to the jet, the observation of such a
structure in the $\RM$s of the optically thin portions of AGN jets is
a robust indicator of the presence of ordered, toroidal magnetic
fields.  However, the strength of this evidence does depend upon the
particulars of the jet model, and more importantly the beam
resolution.

At small inclinations, insufficient radial range may exist to argue
unambiguously for the correlated bilateral morphology.  At large
inclinations, the transverse structure of the jet may remain
unresolved, leading to the previously-mentioned issues surrounding
beam convolution.  This is illustrated in the large-beam $\sgn(\RM)$
maps (Figure \ref{fig:sRM2}, bottom), in which the unresolved maps
show relatively little structure in all cases.  While it still appears
possible to determine if ordered toroidal magnetic fields exist within
the Faraday screen, the large beam limits the amount of asymmetric
structure present when toroidal fields are sub-dominant.  This
difficulty vanishes when the transverse jet structure is
resolved.

The morphology of the $\sgn(\RM)$ maps may also be used to
observationally probe the location of the Faraday screen, ruling out
contributions from distant, randomly oriented foreground clouds.
While $\pc$-scale clouds have been detected via free-free absorption
in a handful of sources \citep{Jone_etal:96,Walk_etal:00}, typically
they are incapable of reproducing the large-scale radial correlations
associated with Faraday rotation within the jet sheath.  This may be argued on
general grounds: Any model for the foreground Faraday screen must have
structure on scales comparable to the jet width in order to mimic the
observed transverse gradients.  However, since in this case the
Faraday screen is not associated with the jet itself, this implies that
there must be structure parallel to the jet as well.  Thus, we do not
expect to see the bilateral morphology associated with the
Faraday rotation due to a jet sheath.

To illustrate this explicitly, in the right-most panels of Figure
\ref{fig:sRM2} we show two examples of random distant foreground
screens, in which $d\RM/d\ell$ is given by a Gaussian random field
with power-law power spectra, $P_k\propto k^{a_r}$.  To emphasize the
diagnostic ability of the $\sgn(\RM)$ map, we subtract off the
flux-weighted average $\RM$, which is degenerate with a Galactic
contribution, in order to ensure that the $\sgn(\RM)$ map exhibits
significant structure.  These realizations are not intended to
represent physical models for the plasma conditions within the Faraday
screen.  Rather, they simply provide two well-defined examples of the
beam-convolved $\sgn(\RM)$ maps associated with foreground screens
that are uncorrelated with the jet itself.  In keeping with this
motivation, we choose a flat spectrum ($a_r=0$), which necessarily
produces considerable structure on angular scales comparable to the
beam, and a red spectrum ($a_r=-3$) which typically generates
structure on the largest relevant scale (e.g., the jet length).

The flat-spectrum foreground screen produces $\sgn(\RM)$ maps that are
similar in structure to those arising when the toroidal magnetic field
within the jet sheath vanishes.  As a consequence, our discussion of
the $\sgn(\RM)$ maps as indicators of the presence of ordered toroidal
fields applies to the distinction between local and distant,
flat-spectrum Faraday screens as well.  That is, the presence of
ordered, bilateral structures in the optically thin portions of the
jet necessarily excludes this kind of distant screen.

The red-spectrum foreground screen is more problematic.
In this case large-scale structures characteristic of red spectra are
apparent, frequently extending substantial distances along the jet
axis.  As a result, distinguishing between red-spectra foregrounds and
nearby Faraday rotation is difficult at large beam widths.  This
situation is ameliorated somewhat by smaller beams, at which not only
the general morphology but also the detailed structure of the
$\sgn(\RM)$ maps are apparent (cf. the canonical and red-spectrum
models at small beams in Figure \ref{fig:sRM2}).  Despite this, if the
jet sheath is efficiently heated, this complication may remain, making
the distinction between foreground and nearby Faraday rotation for any
individual source challenging.

\subsection{Implications for Jet Models}

Because the $\RM$ depends upon the thermal particles that are directly
modeled in jet theory, it provides a more robust handle upon the
parameters of jet models.  Thus, in addition to addressing prior
suggestions about the source of the $\RM$s associated with AGN jets,
we may also confront the observational applicability of our present
understanding of how relativistic jets are produced and propagate.

First among these is the relationship between the nonthermal particles
and the other jet parameters.  By normalizing the accretion rate via
the typical $\RM$ observed ($\sim10^3\,\rad\,\m^{-2}$), we have been
able to determine the efficiency with which the electromagnetic energy
within the jet is dissipated into nonthermal particles
producing the emission.  For our canonical model we find that
$\eta=0.02$ reproduces typical observed fluxes, meaning that the
nonthermal particles constitute approximately $2\%$ of the energy
density within the jet.  This is roughly an order of magnitude larger
than internal energy of the thermal particles within the jet.  Therefore,
when combined with $\RM$ measurements, their luminosity implies that
at $\pc$-scales AGN jets are Poynting-dominated outflows of
nonthermal, relativistic particles.  This is consistent with some
studies of $\kpc$-scale AGN jets, finding that even at those scales
large-scale fields persist and dominate the plasma thermal energy
\citep[see, e.g.,][]{Lape-Kron:05}.

Second, constraining the plasma parameters of the Faraday screen via
$\RM$ profiles is complicated and will generally require careful
modeling of the screen properties, e.g., via simulations like that we
have presented here.  This is because there are significant
degeneracies between a number of otherwise poorly constrained
properties; namely the jet orientation, bulk velocity of the rotating
medium, thermal plasma density and temperature and the strength of the
magnetic field.  As seen in Figure \ref{fig:raysec}, many of these are
varying substantially throughout the screen.  Furthermore, simplifying
assumptions, such as equipartition, can be wrong by orders of
magnitude.  While it may be tempting to calibrate simple, qualitative
models using simulations like the one we present, the relationship
between the circum-jet material and the fast jet core is a strong
function of black hole spin and other model parameters.  Thus, efforts to
quantitatively infer the properties of the circum-jet material will
require additional, self-consistent jet {\em formation} simulations.

Third, we nevertheless find that the ordered toroidal magnetic field
in the jet and its sheath were crucial to producing emission and
Faraday rotation distributions that comport with observations.
This is consistent with simplified, analytical efforts to explain
these properties with a strongly-magnetized jet containing large-scale
ordered toroidally dominated magnetic fields
\citep[see, e.g.][]{lpg05}.
If AGN jets correspond well to GRMHD jet models like that presented
here, observations of the $\RM$ gradients coupled with constraints
upon the jet orientation, particle acceleration and heating rates
within the jet and $\dot{M}$ as estimated by the disk thermal spectra
and luminosity provide a means to estimate the jet power per unit mass accretion rate.
For jets powered via the Blandford-Znajek mechanism, this quantity
is a strong function of the black hole spin \citep{mck05,tnm10},
and hence may provide a means to measure the black hole spin magnitude
for a large number of AGNs.

Fourth, it has been suggested that variations in the polarization angle
across the jet indicate changes in the direction of the intrinsic
magnetic field.  Specifically, in a number of sources (e.g., 3C 272
and 1055+018) the polarization is aligned with the jet near the axis
and becomes orthogonal to the jet at the edges
\citep{att99,push05,zt05,gmjar08}.  Often this is considered evidence for a
toroidally dominated jet core and a poloidally dominated jet sheath.
However, studies of purely toroidal jets have qualitatively reproduced
these polarization distributions as a consequence of projection
effects and relativistic aberration \citep{lpg05,zbb08}.  Seen
explicitly in Figure \ref{fig:exIRMmaps}, our work also qualitatively
reproduces the observed polarization structure without the need to appeal to
variations in the intrinsic magnetic field geometry.

Finally, while the jet simulation is fully 3D, and this
non-axisymmetric structure is clearly evident in the jet flux
distributions, the resulting $\RM$ profiles are remarkably uniform
(where they are sufficiently resolved).  Thus, despite substantial
departures from axisymmetry observed in the jet core, the assumption
of axisymmetry in the surrounding medium seems well-justified.

\section{Conclusions} \label{sec:C}
It has now become possible to quantitatively compare
$\pc$-scale observations of AGN jets
with self-consistent fully 3D numerical
GRMHD simulations of jets.   Due to the need to model the nonthermal particles, direct
comparisons between the emission and polarization properties of
simulated and real jets are not possible, generally requiring
additional post hoc assumptions about the poorly understood
process of particle acceleration within the jet.  In contrast, the
$\RM$s associated with the circum-jet outflows depend only upon the
simulated quantities.  For these reasons AGN jet $\RM$s provide a
potentially more robust diagnostic of the simulation parameters.

Even with relatively low accretion rates,
$\sim10^{-2}\dot{M}_{\rm Edd}$, we are able to reproduce the typical
$\RM$s of $10^3\,\rad\,\m^{-2}$ observed in $\pc$-scale jets.  These
are generally proportional to $\dot{M}^{3/2}$, and thus even moderate
increases in the accretion rate results in substantial increases in
the typical $\RM$s.  As a consequence it is possible to produce the
majority of the observed Faraday rotation via propagation through the
circum-jet material alone.  Were the bulk of the observed $\RM$s to
be produced far from the jet, this would require either $\dot{M}$s
well below that necessary to produce the observed emission or very
efficient dissipation of the magnetic field immediately outside of the
jet (and thus very hot circum-jet plasma).

While well-separated from the emission region, the bulk of the Faraday
rotation occurs in the immediate vicinity of the jet, within the
surrounding jet sheath.  This consists of a highly magnetized,
relativistic outflow, which though moving more slowly than the jet core
is more properly thought of as a continuous extension of the jet
itself than a separate portion of the outflow.  As a consequence,
like the jet core, the magnetic field in the Faraday rotating jet sheath
is dominated by a large-scale, ordered toroidal magnetic field.

The $\RM$ maps associated with our canonical model exhibit well-defined
transverse $\RM$ gradients, associated primarily with the
ordered toroidal magnetic field.  Thus, our results suggest that
efforts to probe the properties of toroidal magnetic fields in the
circum-jet regions are consistent with existing 3D jet formation
simulations, which self-consistently model the fast jet core and the
surrounding Faraday rotating material.  Furthermore, while some
variation in the precise gradients across the jet exist due to
non-axisymmetric jet structures, these are generally small in
comparison to the gradients themselves, implying that $\RM$
observations can produce well-defined estimations of the toroidal
field strength and structure.

The $\RM$ map details are also sensitive to the poloidal
velocities, though in a way that is roughly degenerate with the
dependence upon the mass accretion rate.  However, at $\pc$ distances
the comoving poloidal magnetic fields and toroidal velocities are
sufficiently small that neither produce perturbations in the
transverse $\RM$ profiles that exceed the variations associated with
the non-axisymmetric structure of the jet.  While this is dependent
upon black  hole mass, for reasonable values this implies that efforts
to probe the poloidal magnetic field and toroidal velocities may be
challenging.

Accurately measuring the $\RM$s of AGN jets depends critically upon
the ability to resolve the jet's transverse structure.  This is
generally violated in the radio core, and thus core $\RM$s and
$\RM$-gradients may be typically unreliable.  When the beam is
moderately larger than the jet width finite beam effects result in a
suppression of the $\RM$ relative to the true values, though
well-defined gradients may still exist.  When the beam
is much larger, or when there is significant spectral or
polarization structure on sub-beam scales, the resulting $\RM$s can
be dramatically different from the true values.  This motivates
pushing towards higher frequencies and higher resolutions, as well as
focusing upon resolved regions more than a beamwidth away from the
radio core.  For these purposes, space-based VLBI experiments like
\VSOPII~are likely to be particularly useful.

The large-scale correlations along the jet indicative of ordered
toroidal fields are present even in the $\sgn(\RM)$ maps, producing a
distinctive, bilateral morphology.  Models which invoke Faraday
rotation far from the jet generally have difficulty reproducing this
structure, and may therefore be excluded by $\sgn(\RM)$ measurements.
Despite the fact that sufficiently red fluctuations in the foreground
$\RM$s can mimic the long-range order indicative of toroidal fields,
sufficiently resolved $\RM$ observations may robustly verify the
presence of Faraday rotation in the vicinity of the jet.

While we have made an attempt to explore the various parameters
critical to computing the $\RM$s of simulated jets, by no means has
our investigation been exhaustive.  The most significant limitation is
the restriction to a single time step of a single jet simulation
with a single prescription for the nonthermal particles.  As a
consequence we are not able to address the uncertainties in the $\RM$s
associated with the substantial variability in the accretion onto the
black hole.  More importantly, we cannot address how interactions with
the surrounding medium, known to be important in many cases, will
affect our results.  Where these interactions significantly alter the
properties of the jet they should also make substantial changes to the
particulars of the $\RM$ maps.  However, since the general structure of
the $\RM$s arise from a direct relationship with the order inherent
in the jet's magnetic field, we would expect the generic morphology of
the $\RM$s to persist as long as the jet is toroidally dominated.

\vspace{1cm}
We thank Denise Gabuzda, Larwence Rudnick, Gregory Taylor and Philipp
Kronberg for helpful comments and suggestions.  This work was
supported in part by NASA Chandra Fellowship PF7-80048 (JCM).

\appendix

\section{Relativistic Aberration} \label{app:RA}
In Section \ref{sec:CM:PRTiBF} we described how we performed the
polarized radiative transfer through relativistically moving bulk
flows.  Here we present how the polarization direction changes due to
relativistic aberration, and thus specifying the orientation of the
parallely propagated orthonormal tetrad used to define the Stokes
parameters.  We will make use of four-vectors and covariant analysis,
though the result will be quoted in terms of the three vectors we have
used to define the relevant plasma quantities.

We choose to orient the Stokes parameters so that the projected,
positive $z$-axis is ``up'' in the image plane.  This is necessarily
orthogonal to the wave four-vector, $k^a=(1,\bmath{\hat{k}})$, and
thus,
\begin{equation}
e^a = (0,\bmath{e}) - C k^a\,,\quad
\bmath{e}\equiv\bmath{\hat{z}}-\left(\bmath{\hat{z}}\cdot\bmath{\hat{k}}\right)\bmath{\hat{k}}\,,
\end{equation}
in which $C$ is an arbitrary scalar.  Generally, $e^a$ is only defined
up to $C$, corresponding to the ambiguity associated with the
time-component of the polarization four-vector in any given
frame\footnote{Alternatively, this may be thought of as the gauge
  freedom associated with the four-vector potential.}.  Thus we are
free to require $e^a$ be purely space-like in the jet frame, i.e., with
$u^a=(\Gamma,\Gamma\bmath{\beta})$, we set $C$ such that $u^ae_a=0$,
obtaining,
\begin{equation}
C = \frac{\bmath{\beta}\cdot\bmath{e}}{1-\bmath{\beta}\cdot\bmath{\hat{k}}}
\end{equation}
This now defines the ``up'' direction, as seen in the jet frame.

In the emitter's frame, the natural orientation for the Stokes
parameters is aligned with the projected magnetic field, which is
given by
$\bmath{b}'-(\bmath{b}'\cdot\bmath{\hat{k}}')\bmath{\hat{k}}'$, where all primed
quantities are in the jet frame (though note that $\hat{k}'$ is the
spatial components of $k^a$ in the jet frame, and thus is not
unit-normalized).  Thus,
\begin{equation}
\begin{aligned}
\cos\xi
&=
\frac{
  \bmath{e}'\cdot\left[\bmath{b}'-(\bmath{b}'\cdot\bmath{\hat{k}}')\bmath{\hat{k}}'\right]
}{
  |\bmath{e}'||\bmath{b}'-(\bmath{b}'\cdot\bmath{\hat{k}}')\bmath{\hat{k}}'|
}\\
&=
\frac{
  \bmath{e}'\cdot\bmath{b}'
}{
  \sqrt{e^a e_a \left[ b^2-(b^ak_a/u^ak_a)^2\right]}
}\\
&=
\frac{
  e^a b_a
}{
  \sqrt{e^a e_a \left[ b^2-(b^ak_a/u^ak_a)^2\right]}
}\,,
\end{aligned}
\end{equation}
where we repeatedly used the fact that $e^a$ and $b^a$ are purely
space-like in the jet frame.  Via explicit computation, we have in the
lab frame
\begin{equation}
\begin{aligned}
e^a e_a &= \bmath{e}\cdot\bmath{e} = 1 -
\left(\bmath{\hat{z}}\cdot\bmath{\hat{k}}\right)^2\\
e^a b_a &= \bmath{e}\cdot\bmath{b} - \frac{\Gamma\bmath{\beta}\cdot\bmath{e}}{u^a k_a} k^b b_b\\
&= \bmath{e}\cdot\bmath{b} + \Gamma\bmath{\beta}\cdot\bmath{e} b \cos\vartheta\\
\frac{b^a k_a}{u^b k_b}
&= -b\cos\vartheta\,.
\end{aligned}
\end{equation}
Therefore,
\begin{equation}
\cos\xi = \frac{
\bmath{e}\cdot\left(\bmath{b} + \Gamma\bmath{\beta} b\cos\vartheta\right)
}{
\sqrt{
1-\left(\bmath{\hat{z}}\cdot\bmath{\hat{k}}\right)^2
}
b\sin\vartheta
}\,,
\end{equation}
which reduces to Equation (\ref{eq:polabb}) upon noting that
$\bmath{e}=\left(\bmath{1}-\bmath{\hat{k}}\bmath{\hat{k}}\right)\cdot\bmath{\hat{z}}$

\bibliography{jetrm}
\bibliographystyle{apj}

\end{document}